\documentclass[]{elsarticle}

\usepackage{lineno,hyperref}
\usepackage{amsmath}
\usepackage{amsfonts}
\usepackage{amssymb}
\modulolinenumbers[5]
\usepackage{subfigure}
\usepackage{xcolor}
\usepackage{placeins}
\usepackage{caption}
\usepackage{dcolumn}
\newcolumntype{.}{D{.}{.}{2.1}}

\journal{Journal}

\graphicspath{{figs/}}

\begin{document}

\begin{frontmatter}

\title{Proteinoid microspheres as proto-neural networks}

%% Authors and affiliations

\author{Panagiotis Mougkogiannis*}
\author{Andrew Adamatzky}
\address{Unconventional Computing Laboratory, UWE, Bristol, UK}

\cortext[cor1]{Corresponding author: Panagiotis.Mougkogiannis@uwe.ac.uk (Panagiotis Mougkogiannis)}

\begin{abstract}
Proteinoids, also known as thermal proteins, possess a fascinating ability to generate microspheres that exhibit electrical spikes resembling the action potentials of neurons. These spiking microspheres, referred to as protoneurons, hold the potential to assemble into proto-nano-brains. In our study, we investigate the feasibility of utilizing a promising electrochemical technique called differential pulse voltammetry (DPV) to interface with proteinoid nano-brains. We evaluate DPV's suitability by examining critical parameters such as selectivity, sensitivity, and linearity of the electrochemical responses. The research systematically explores the influence of various operational factors, including pulse width, pulse amplitude, scan rate, and scan time. Encouragingly, our findings indicate that DPV exhibits significant potential as an efficient electrochemical interface for proteinoid nano-brains. This technology opens up new avenues for developing artificial neural networks with broad applications across diverse fields of research.
\end{abstract}


\end{frontmatter}
\section{Introduction}

While there are numerous prototypes of organic electronic devices \cite{ji2019recent,fahlman2019interfaces,mao2019bio,matsui2019flexible}, very few, if any, demonstrate substantial degrees of stability or bio-compatibility \cite{feron2018organic}. This is why we propose to explore thermal proteins \cite{fox1992thermal}, a unique class of organic devices, as a substrate and architecture for future non-silicon massive parallel computers.  Proteinoids, also known as thermal proteins, are derived by subjecting amino acids to high temperatures until they reach their melting point, leading to polymerization and the formation of polymeric chains~\cite{fox1992thermal}. This polymerization process occurs in the absence of solvents, initiators, or catalysts, under an inert atmosphere, typically at temperatures ranging from 160 to 200 °C. Specifically, the tri-functional amino acids, such as glutamic acid, aspartic acid, or lysine, undergo cyclisation at elevated temperatures and serve as solvents and initiators for the polymerization of other amino acids \cite{harada1958thermal,fox1992thermal}. The intriguing capacity of proteinoid microspheres to generate action-potential-like spikes and spike trains has led to their consideration as analogs of proto-neurons, which are neuron-like cells that function without metabolic processes \cite{fox1995experimental,rizzotti1998did,matsuno2012molecular}.

We explore the concept of nano brains (PNBs)~\cite{baluvska2021biomolecular,callaway2021deepmind} to evaluate feasibility of proteinoid microspheres for physical imitation of  artificial neuron networks (ANNs)~\cite{2018Artificial}. Namely, we aim to imitate neuronal responses to external stimuli~\cite{zhang2019artificial,nwadiugwu2020neural,fox1982updated,goi2020perspective,fiers2013nanophotonic} in PNBs. We use  differential pulse voltammetry (DPV) for assessing capabilitie of PBNs for pattern recognition.  Comparative analyses of DPV and alternative approaches are conducted, and the current research status in this field is reviewed. The implications of the findings for subsequent research are also discussed.

In this study, we  employ techniques from the fields of electrochemical neuroscience, artificial neural networks (ANNs), and pattern recognition to analyze the spikes generated by PNBs. Specifically, we  utilize differential pulse voltammetry (DPV) to measure the electrical signals produced by PNBs. Through this analysis, we aim to understand the behavior of PNBs and evaluate their capability for pattern recognition. Our objective is to review existing literature that explores the relationship between PNBs and ANNs, identify any gaps or limitations in the current understanding, and propose a research methodology to investigate the potential of PNBs in the field of pattern recognition~\cite{kwon2016neuron, ramachandran2017theranostic,mougkogiannis2023low}.

Now, let's delve into the examination of the neural networking capabilities of biological neurons. A group of researchers at NIST (National Institute of Standards and Technology) has made significant advancements in this area by developing an artificial neuron that exhibits an astonishing firing rate of 100 billion times per second~~\cite{2018Artificial}. This remarkable speed surpasses the firing rate of a human brain cell by approximately tenfold. The research article highlights the use of niobium nitride, a superconducting material, in the artificial neuron. This material allows the neuron to switch between two distinct electrical resistance states when exposed to magnetic fields. The article discusses the possibilities and challenges associated with creating "neuromorphic" hardware that emulates the complex functioning of the human brain~\cite{2018Artificial}.

In their research, Wan et al. presented a breakthrough in the field of artificial neurons by showcasing the functionality of an artificial sensory neuron capable of gathering optic and pressure data from photodetectors and pressure sensors, respectively~\cite{kwon2016neuron}. This  neuron can transmit the combined information through an ionic cable and integrate it into post-synaptic currents using a synaptic transistor. The study highlights the significance of synchronizing two sensory cues, as it activates the artificial sensory neuron at different levels, enabling the control of skeletal myotubes and a robotic hand. Furthermore, the research demonstrates that the artificial sensory neuron enhances recognition capabilities for combined visual and haptic cues through the simulation of a multi-transparency pattern recognition task~\cite{kwon2016neuron}.

In their study, Boddy et al. employ artificial neural networks (ANNs) to effectively identify and classify marine phytoplankton using flow cytometry data, showcasing the capability of ANNs in recognizing patterns in biological data~\cite{ramachandran2017theranostic}. The article provides an overview of the structure and training process of three types of ANNs: backpropagation (multilayer perceptron), radial basis function (RBF), and learning vector quantization. These ANNs utilize supervised learning techniques and are well-suited for biological identification purposes. Additionally, the study highlights the effectiveness of Kohonen self-organizing maps (SOM) and adaptive resonance theory (ART) as classification methods~\cite{ramachandran2017theranostic}.

In their research, Syed and colleagues introduce a groundbreaking concept that goes beyond the traditional fixed feedforward operation commonly found in contemporary artificial neural networks~\cite{syed2023atomically}. The study presents a novel class of synthetic neurons capable of adapting their functionality in response to feedback signals from neighboring neurons. These synthetic neurons demonstrate the ability to emulate complex brain functions, including spike frequency adaptation, spike timing-dependent plasticity, short-term memory, and chaotic dynamics~\cite{syed2023atomically}.

Baluska et al. explore the evolutionary perspective of biomolecular structures and processes that contribute to the emergence and maintenance of cellular consciousness~\cite{baluvska2021biomolecular}. The proposition suggests that subcellular components, such as actin cytoskeletons and membranes, play a crucial role in nano-intentionality. This is attributed to the inherent structural adaptability of individual biomolecules, extending beyond cellular boundaries~\cite{baluvska2021biomolecular}.

Present paper focuses on exploring the capabilities of proteinoid nano brains (PNBs) in processing signals obtained from a differential pulse voltammetry (DPV) electrode and their potential for pattern recognition, drawing inspiration from artificial neural networks (ANNs)~\cite{wan2020artificial,yegnanarayana1994artificial,syed2023atomically}. The objective of this study is to investigate the ability of PNBs to detect spikes induced by DPV signals. We aim to assess the responsiveness of PNBs to DPV signals and their capacity to generate ANNs for pattern recognition purposes. Experimental results are presented, evaluating the pattern recognition performance of PNBs using DPV signals. The paper concludes by discussing the implications of the findings and providing recommendations for future research.

\section{Methods}

\begin{figure}[!tbp]
\centering
\includegraphics[width=1\textwidth]{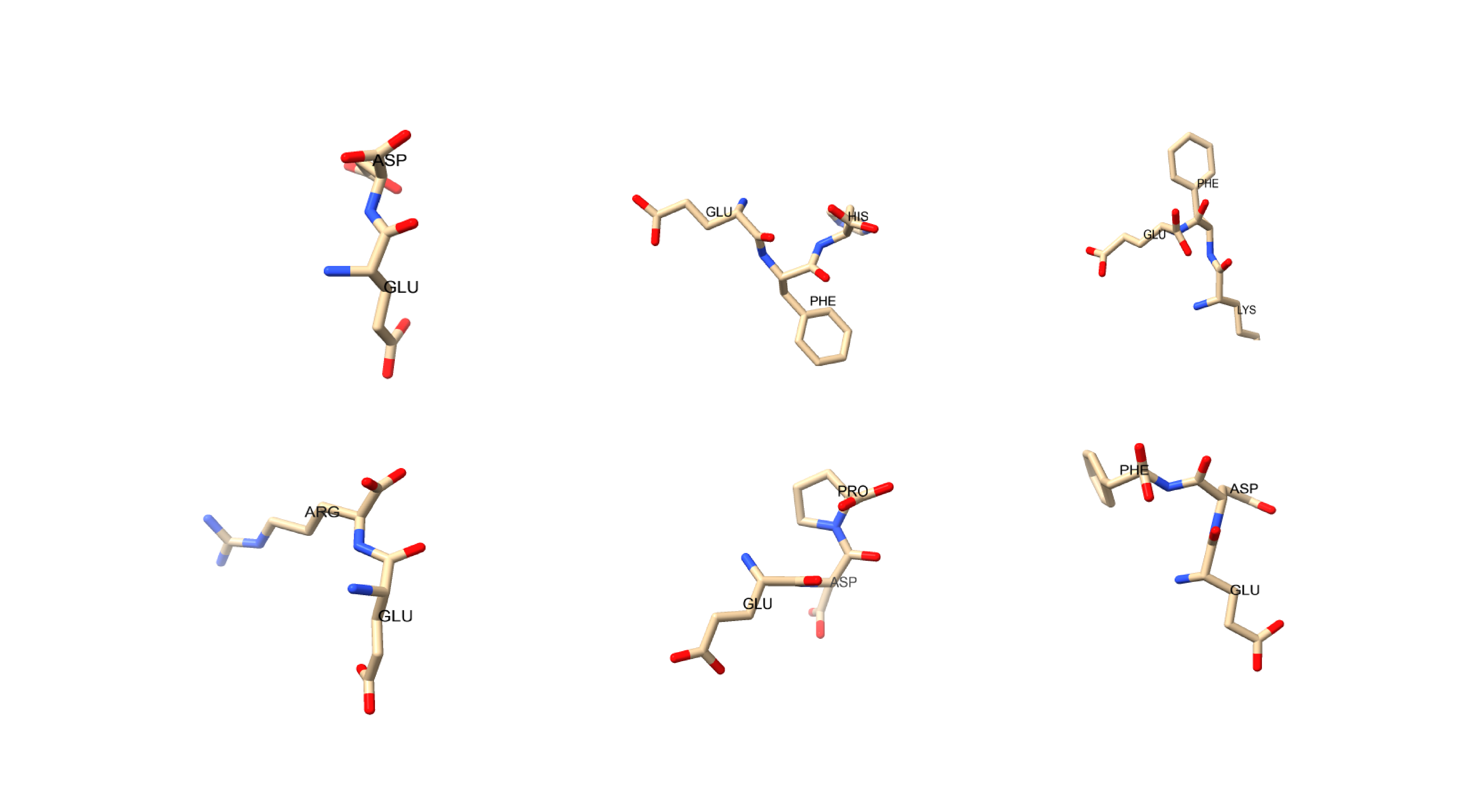}
\caption{Chemical structures of L-Glu:L-Asp, L-Glu:L-Phe:L-His, L-Lys:L-Phe:L-Glu, L-Arg,L-Glu, L-Glu:L-Asp-L-Pro, L-Glu:L-Asp:L-Phe.
}
\label{jn gsdbzdbdbd bk bkb k}
\end{figure}

High-purity amino acids, including L-Phenylalanine, L-Aspartic acid, L-Histidine, L-Glutamic acid, and L-Lysine (Fig.~\ref{jn gsdbzdbdbd bk bkb k}), were acquired from Sigma Aldrich with a purity exceeding 98\%. The synthesis of proteinoids followed previously established methods~\cite{mougkogiannis2023transfer}. The structural analysis of the proteinoids was conducted using scanning electron microscopy (SEM) with the FEI Quanta 650 equipment. Characterization of the proteinoids was performed using FT-IR spectroscopy~\cite{mougkogiannis2023transfer}.

To measure the electrical activity of the proteinoids, iridium-coated stainless steel sub-dermal needle electrodes (Spes Medica S.r.l., Italy), and high-resolution data logger equipped with a 24-bit A/D converter (ADC-24, Pico Technology, UK) were used. The electrodes were configured in pairs to measure the potential difference between them, with an inter-electrode distance of approximately 10~mm. Electrical activity was recorded at a sampling rate of one sample per second. The data logger recorded multiple measurements (typically up to 600 per second) and stored the mean value for analysis.

Differential pulse voltammetry (DPV) can be used to take accurate measurements with the Zimmer \& Peacock Anapot EIS. The Anapot EIS provides users with the flexibility to define measurement parameters for conducting differential pulse voltammetry (DPV) experiments. In order to perform a DPV measurement, several key parameters need to be specified, as follows the equilibrium time is set to 100~sec, the potential scan starts at -8~V, the potential scan ends at 8~V, the potential step size is set to 0.001~V, the pulse amplitude is set to 0.2~V, the pulse width is specified as 0.08~sec, the scan rate is set to 0.001~V/s.
During the measurement process, the Anapot EIS applies brief pulses to the working electrode in small steps. It measures the current response twice in each step, capturing the current values before and after the pulse. This process is repeated until every phase of the potential scan is completed.

By precisely controlling the measurement parameters and acquiring current response data at different potentials, the Anapot EIS enables comprehensive analysis and characterization of samples through differential pulse voltammetry.

\section{Results}

\begin{figure}[!tbp]
\centering
\includegraphics[width=1\textwidth]{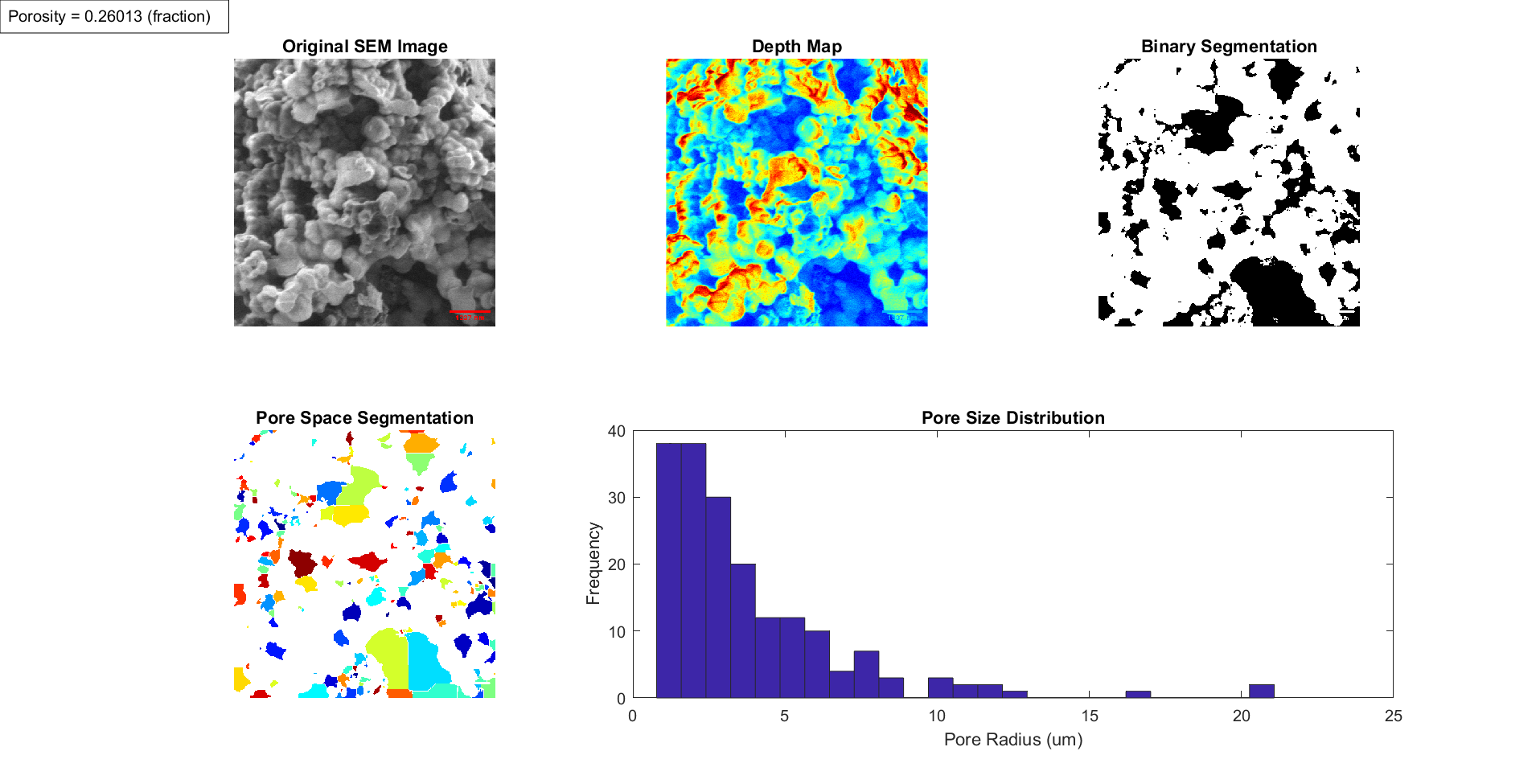}
\caption{Porosity of proteinoids. This graph shows average porosity (3.7982 um) of proteinoids, represented by a depth map, binary segmentation, pore space segmentation, and pore size distribution in um
\cite{rabbani2017dynamic}.}
\label{jn eegbebb}
\end{figure}

Scanning electron microscopy revealed the complex proteinoid molecular network tuned to 1337 nm porosity (Fig.~\ref{jn eegbebb}). The network of molecules observed in our experiments appears to show some morphological similarity to neural cultures~\cite{kon2000information}.  

\begin{figure}[!tbp]
\centering
\includegraphics[width=1\textwidth]{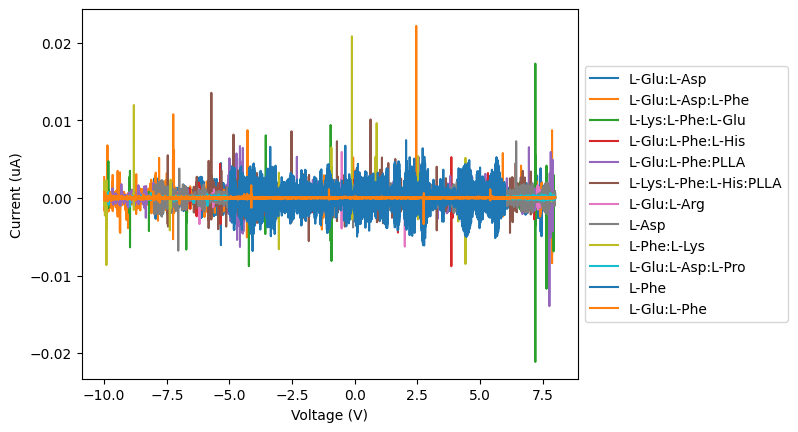}
\caption{DPV measurements of 12 different proteinoids. The heigt of each peak is proportional to the number of microspheres present in the sample.
}
\label{ppsoiunpsssazx999}
\end{figure}

The results of Figure~\ref{ppsoiunpsssazx999} suggests that proteinoids behave as electrical semi-conductors, likely due to their amino acid chain structure. This type of electrical activity is reminiscent of that observed in neurons within the nervous system~\cite{HedgesElectrical,MIT2018Seeing,xu2018collective,pinotsis2023cytoelectric}. Although the level of electrical activity is lower than that of neurons, the similarity in its behaviour is intriguing. Despite this, the results of Figure~\ref{ppsoiunpsssazx999} suggest that the electrical nature of proteinoids may exhibit properties similar to what is observed in biological cells~\cite{ouldali2020electrical,rosenberg1962electrical,mougkogiannis2023low}. The data from Fig.~\ref{ppsoiunpsssazx999} is further supported by our previous research that proteinoids could synchronize electrical activity~\cite{mougkogiannis2023transfer,mougkogiannis2023low}. 

Based on the data presented in Table~\ref{vdbfngdwfwvwvq;}, it can be observed that the proteinoids exhibited a considerable range in the quantity of spikes they generated. The presence of this phenomenon was demonstrated through the fluctuating quantity of spikes that were detected in the voltage-current graphs. The findings indicate that the mean number of spikes observed was 385.8, with a range spanning from 8 spikes for the L-Glu:L-Asp:L-Pro sample to 900 spikes for the L-Phe sample.  The time duration metrics of the previously mentioned spikes varied from 20.71 seconds for L-Phe to 2541 seconds for L-Glu:L-Asp:L-Pro. The data indicates that the proteinoids exhibited a mean inter-spike interval of 425.30 seconds. 

The combination of L-Glu, L-Asp, and L-Pro in a ratio of L-Glu:L-Asp:L-Pro resulted in a lower number of spikes, 8 compared to other combinations. Additionally, the mean inter spike interval (2541.00 sec) for these three combinations was slightly higher than that of the other combinations.

\begin{figure}[!tbp]
\centering
\includegraphics[width=1\textwidth]{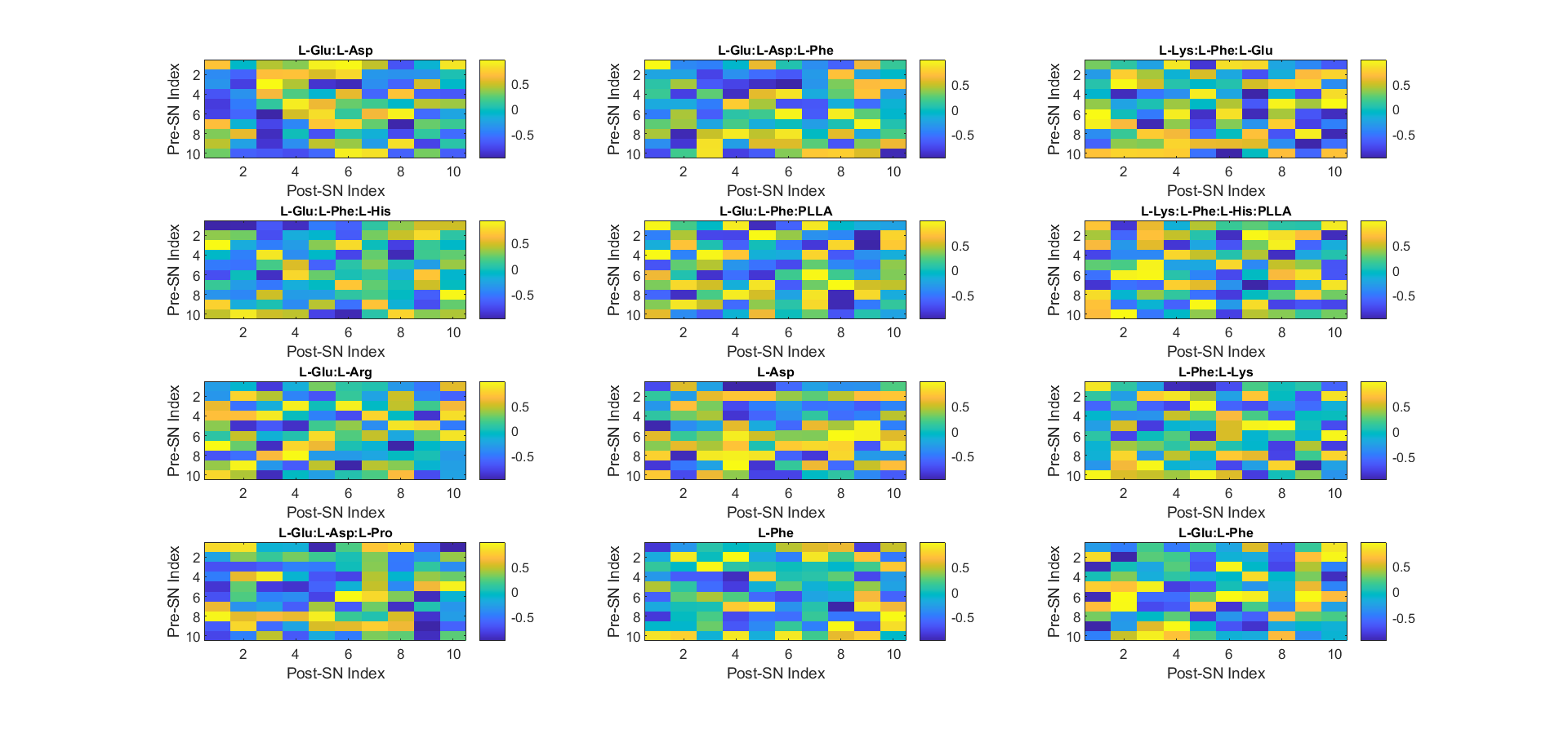}
\caption{The presented colour map depicts the PSI and PPI values of various proteinoids. PSI, or post-synaptic index, quantifies the chemical or functional potency of inter-neuronal connections within a network. PPI stands for post postsynaptic index. It quantifies the efficacy of inter-neuronal connections in a given network.
Darker colours of blue indicate elevated PSI values, whereas lighter colours of green indicate elevated PPI values. The map illustrates the correlation between post-synaptic and pre-synaptic neurons and their influence on proteinoid function.}
\label{sdgfahdsgndhmj,,}
\end{figure}

Figure~\ref{sdgfahdsgndhmj,,} depicts the relationship between pre-synaptic and post-synaptic neurons and their impact on proteinoid activity. The terms presynaptic and postsynaptic refer to the two sides of a synapse, which is the point of connection between two neurons or between a neuron and a target cell. The presynaptic neuron releases neurotransmitters, which are chemical messengers facilitating intercellular communication. The postsynaptic neuron receives and responds to the neurotransmitter by either firing or not firing an action potential~\cite{gurth2020synaptic}.

The code in Fig.~\ref{sdgfahdsgndhmj,,} assumes that proteinoid samples can form microspheres that function as primitive neuronal analogues, and that the measured potential values correspond to the electrical activity of these microspheres.
Moreover, the code utilises temporal coding to convert current values (in $\mu$A) to binary values, indicating the presence or absence of a spike at a particular time for each proteinoid microsphere. A spike denotes an abrupt rise in the electrical potential across the membrane of a neuron or a proteinoid microsphere, signifying its activation or firing. The ANN assigns 10 proteinoid microspheres to 10 neurons and employs synaptic weights to depict their interconnections. Synaptic weights represent the relative strength or influence of one neuron on another. The synaptic weights are initialised randomly, without of any pre-existing knowledge or learning. Fig.~\ref{sdgfahdsgndhmj,,} employs post-synaptic and pre-synaptic indices for neuron labelling in the ANN. The post-synaptic neuron receives signals from another neuron, while the pre-synaptic neuron sends signals to another neuron. The algorithm employs heatmaps to depict the synaptic weights of individual samples. Each cell in the heatmap represents a combination of post-synaptic and pre-synaptic neurons, and the colour denotes the synaptic weight value.

\begin{table}
\caption{Initial values of W matrix for a temporal coding neural network with 10 neurons and random weights.}
\centering
\begin{tabular}{|*{10}{.|}}
\hline
-1.0 & 1.0 & 1.0 & 1.0 & -1.0 & 1.0 & -1.0 & 1.0 & -1.0 & 1.0 \\
\hline
1.0 & -1.0 & 1.0 & 1.0 & 1.0 & -1.0 & 1.0 & 1.0 & -1.0 & 1.0 \\
\hline
1.0 & -1.0 & -0.4 & -1.0 & -0.8 & -1.0 & -1.0 & 1.0 & -1.0 & -1.0 \\
\hline
-1.0 & -1.0 & -1.0 & -1.0 & -1.0 & -1.0 & -1.0 & -1.0 & 1.0 & 1.0 \\
\hline
-1.0 & -1.0 & 0.2 & -1.0 & -1.0 & -1.0 & 1.0 & -1.0 & 1.0 & 1.0 \\
\hline
-1.0 & 1.0 & 0.5 & 1.0 & 1.0 & 1.0 & -1.0 & 1.0 & -1.0 & 0.7 \\
\hline
1.0 & -0.5 & -0.6 & 0.7 & 1.0 & 1.0 & -0.7 & 1.0 & 1.0 & -1.0 \\
\hline
-1.0 & 1.0 & 1.0 & -0.9 & -1.0 & -1.0 & 1.0 & -1.0 & 1.0 & 1.0 \\
\hline
1.0 & -1.0 & 1.0 & 1.0 & -1.0 & -1.0 & 1.0 & -1.0 & 1.0 & -1.0 \\
\hline
0.3 & 1.0 & -1.0 & -1.0 & -0.2 & -1.0 & 0.1 & -1.0 & 1.0 & -1.0 \\
\hline

\end{tabular}
\label{jkbjkbbjoooh}
\end{table}

 Initially, the synaptic weights of a 10-neuron network were randomly initialised within the range of -1 to 1, as presented in Table \ref{jkbjkbbjoooh}.  Furthermore, the way in which the input is converted into an output by the activation function (as depicted in Figure~\ref{sdgfahdsgndhljnkjjmj,,}) may be understood as a depiction of the information exchange among neurons in the nervous system, wherein a greater input would yield a correspondingly higher output. Furthermore, the temporal codes may be regarded as a depiction of the action potential within the nervous system, in which the potential must attain a threshold prior to the enhancing of temporal codes and the consequent activation of output. 

Proteinoids can be used to make interpretations and analogies of the nervous system based on this mathematical relationship. 

Let  $t \in \mathbb{R}^n$ be a vector of time values in seconds, $p \in \mathbb{R}^{n \times 12}$ be a matrix of potential values in volts for 12 samples, $N \in \mathbb{N}$ be the number of neurons in the ANN, 
$T \in \mathbb{R}^+$ be the time window for temporal coding in seconds, $\theta \in \mathbb{R}^+$ be the threshold for spike detection in volts, $c \in \{0,1\}^{N \times n}$ be a matrix of temporal codes for each neuron over time, $W \in [-1,1]^{N \times N}$ be a matrix of synaptic weights between neurons. Then, for each sample  $i = 1,\dots,12$ we have 
where $1_{[p_{j,i} > \theta]}(j)$ is an indicator function that returns 1 if $p_{j,i}$ is greater than $\theta$, and 0 otherwise.

Neurons are distinguished by their unique temporal coding, input parameters, and synaptic weights. When the temporal code ($c_{:,j}$) exceeds the threshold parameter ($\theta$), the neuron representation fires an action potential through its axonal connections, similar to a real neuron. The proteinoid neurons bear resemblance to the neurons in an actual nervous system. The synaptic weights ($W$) in a neural network are similar to the synapses in a biological nervous system, as they govern the potency of the link between axons and dendrites.

Figures~\ref{ppsoiunpsssazx999} and ~\ref{sdgfahdsgndhljnkjjmj,,} differ in stimulation level. Figure ~\ref{ppsoiunpsssazx999} utilises DPV, whereas Figure ~\ref{sdgfahdsgndhljnkjjmj,,} employs a power source that supplies a stable voltage through the proteinoid solution.
According to the findings of the current research, proteinoids are capable of interpreting and responding to various forms of stimulation. When stimulated with DPV (Figure~\ref{ppsoiunpsssazx999}), the proteinoid solution unexpectedly produced oscillating signals as if it were a nervous system analogue. Again unexpectedly, when the proteinoid solution was stimulated with a stable power source (Figure~\ref{sdgfahdsgndhljnkjjmj,,}), discrete signals were produced.
Similar to a nervous system, the proteinoid solution was able to interpret and respond to various forms of stimulation, as indicated by the results. In addition, it appears that the stability of the power source may affect the modulation of the response. This phenomenon can be attributed to the proteinoids' ability to distinguish between the DPV and the stabilised power source, and to react accordingly.

%\begin{figure}[!tbp]
%\centering
%\includegraphics[width=1\textwidth]{figs/sos3.png}
%\includegraphics[width=1\textwidth]{Reponse2stimulation.pdf}
%\caption{At the nanoscopic level, an ensemble of proteinoid microspheres reveals an intricate and complex network of molecule, each finally tuned to 1337 nm. The proteinoid sample exhibits the following characteristics as determined by Matlab: contrast = 0.1928, correlation = 0.9608, energy = 0.1229, and homogeneity = 0.9047.
%\}
%\label{proteinoidmicroscopic}
%\end{figure}

%\begin{figure}[!tbp]
%\centering
%\includegraphics[width=1\textwidth]{figs/DPV_all.png}
%\includegraphics[width=1\textwidth]{Reponse2stimulation.pdf}
%\caption{The Differential Pulse Voltammetry (DPV) of 12 different proteinoid molecules showed unique spiking profiles.
%}
%\label{jn eegbeb bk bkb k}
%\end{figure}

\begin{table}[!tbp]
\centering
\caption{Proteinoid spike characteristics: Number of spikes, Mean inter spike intervals (sec), and frequency of spiking (mHz). The series of measurements obtained for these proteinoids  showed a threshold of spiking at 0.0005 uA with a minimum peak-distance of 5 sec.}
\begin{tabular}{|c|c|c|c|}
	\hline
	Proteinoid  & Number  & Mean  & Frequency      \\
      &  of  &  Inter-Spike  &  of     \\
   & Spikes   & Interval (s) & Spiking (mHz)      \\
	\hline\hline
	
	L-Glu:L-Asp & 726 & 22.24 &  44.97    \\
 L-Glu:L-Asp:L-Phe & 359 & 50.48 & 19.80   \\
 L-Lys:L-Phe:L-Glu & 210 & 85.75 &   11.66   \\
 L-Glu:L-Phe:L-His & 382 & 42.21 & 23.69      \\
  L-Glu:L-Phe:PLLA & 555 & 32.71 &  30.57\\
   L-Lys:L-Phe:L-His:PLLA & 195 &  77.29 & 12.94\\
    L-Glu:L-Arg & 29 & 544.68 & 48.28\\
     L-Asp & 779 & 20.71 & 36.26\\
      L-Phe:L-Lys & 28 & 666.11& 1.50\\
       L-Glu:L-Asp:L-Pro & 8 & 2541.00 & 0.39\\
        L-Phe & 900 & 12.32 & 81.15\\
         L-Glu:L-Phe & 12 & 1412.55 & 0.71\\
 
	\hline
	\end{tabular}
\label{vdbfngdwfwvwvq;}
\end{table}

%\begin{table}[!tbp]
%\centering
%\caption{
%}
%\begin{tabular}{|c|c|c|c|}
%	\hline
%	Proteinoid  & $\mu$ & $\sigma$ &       \\
 %      &  Mean  &  Std  & NLL      \\
  %  &  &  &     \\
%	\hline\hline
	
%	L-Glu:L-Phe:L-His& 3247.9 & 760.83  & 148.06   \\
% L-Glu:L-Phe & 3534.3 & 453.94 &  272.21   \\
% L-Phe:L-Lys & 3742.9  & 517.55 &   248.54    \\
% L-Phe & 3400.8 & 1144.8   & 122.58      \\
% L-Asp & 2237.4 & 745.87  & 118.03      \\
%	\hline
%	\end{tabular}
%\label{dnckscks}
%\end{table}

%\begin{figure}[!tbp]
%\centering
%\includegraphics[width=1\textwidth]{figs/spiketraining.png}
%\includegraphics[width=1\textwidth]{Reponse2stimulation.pdf}
%\caption{This figure shows the spike train plots of 12 different proteinoids, depicting the timing of their activation patterns.
%}
%\label{kfmdklngkbsbb}
%\end{figure}

%\begin{figure}[!tbp]
%\centering
%\includegraphics[width=1\textwidth]{figs/histograms.png}
%\includegraphics[width=1\textwidth]{Reponse2stimulation.pdf}
%\caption{The figure shows the histograms of interspike intervals of 12 different proteinoids using DPV. The mean interspike intervals varied considerably among the different proteinoids, with the shortest mean occuring at 0.000992 msec.
%}
%\label{asdvsabvdbfdbd}
%\end{figure}

\begin{figure}[!tbp]
\centering
\includegraphics[width=1\textwidth]{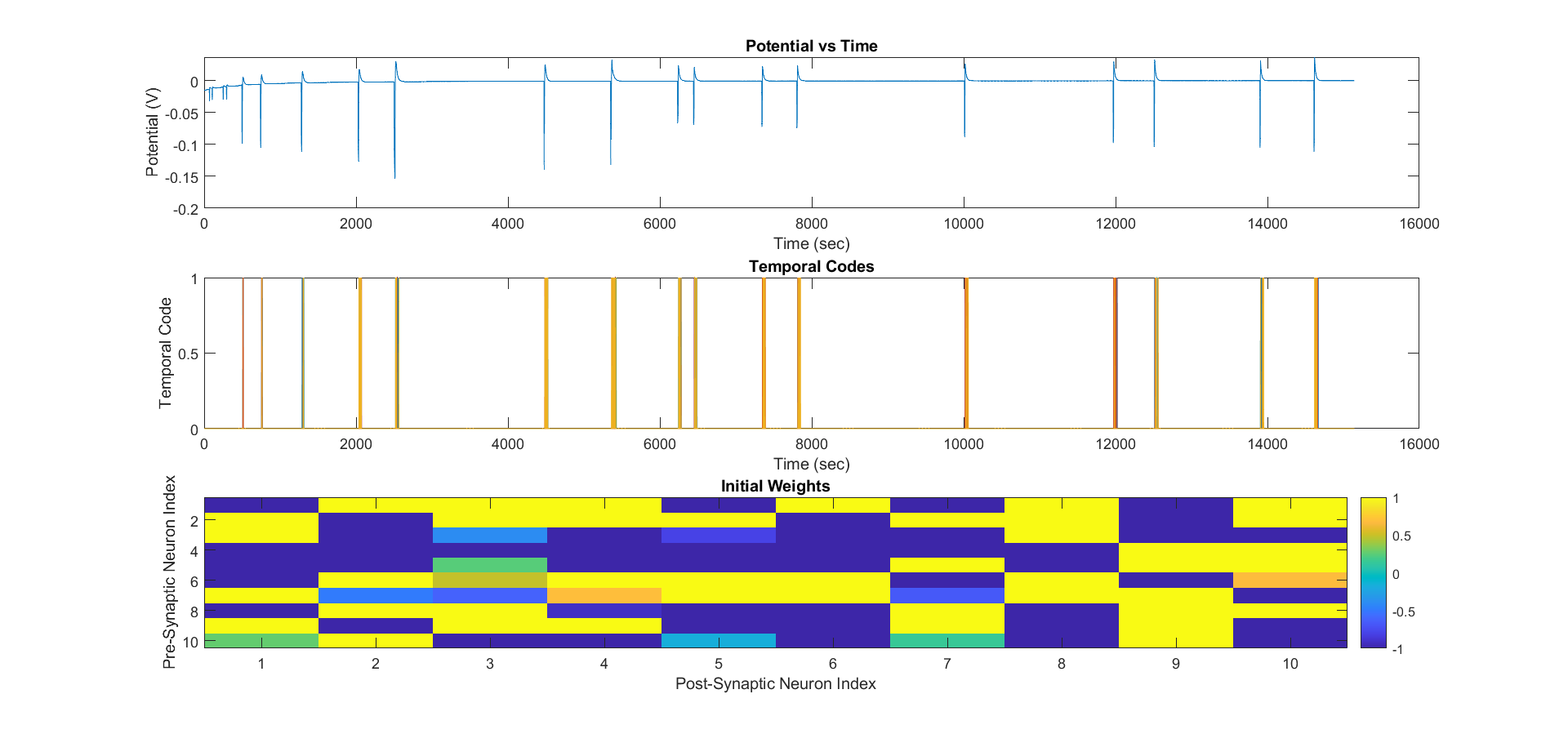}
\caption{The potential of the proteinoid L-Glu:L-Phe when electrically stimulated reveals a temporal code that can be seen in the plots of initial weight and pre- and post-synaptic indices over time.}
\label{sdgfahdsgndhljnkjjmj,,}
\end{figure}

\begin{figure}[!tbp]
\centering
\includegraphics[width=0.9\textwidth]{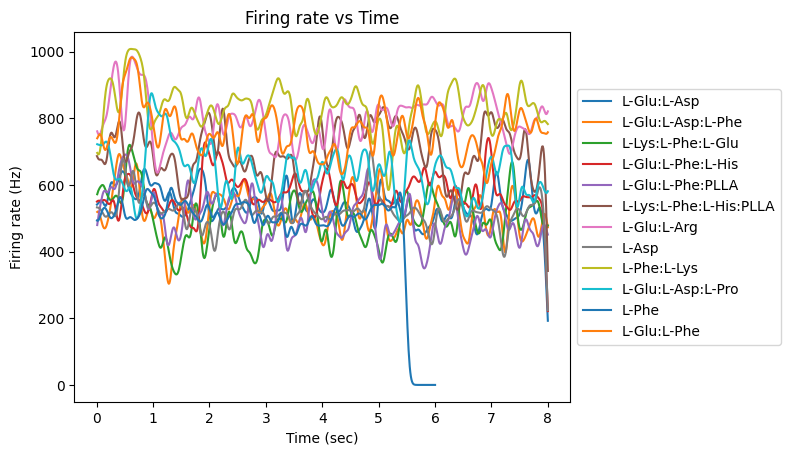}
\caption{L-Phe:L-Lys exhibited the highest firing rate at 1008.2875 Hz.L-Phenylalanine exhibits the lowest firing rate at 591.6153 Hz. The firing rates (in Hz) for the proteinoids 'L-Glu:L-Asp', 'L-Glu:L-Asp:L-Phe', 'L-Lys:L-Phe:L-Glu', 'L-Glu:L-Phe:L-His', 'L-Glu:L-Phe:PLLA', 'L-Lys:L-Phe:L-His:PLLA', 'L-Glu:L-Arg', 'L-Asp', 'L-Phe:L-Lys', 'L-Glu:L-Asp:L-Pro', 'L-Phe', and 'L-Glu:L-Phe' are 535.4877, 436.2721, 542.9443, 567.0562, 498.2888, 650.4798, 732.9516, 529.072, 768.2345, 617.3223, 491.5065, and 665.2995, respectively.
 }
\label{bhjglbjbjbjk,,}
\end{figure}

The results of this study indicate that proteinoid microspheres demonstrate an association between molecular properties and firing rates as presented in Figure~\ref{bhjglbjbjbjk,,}. The firing rate increases significantly with increases in molecular weight and peptide length. This correlation between structural parameters and electrical activity alludes to the possibility of proteinoid microspheres acting as analogs of neurons and forming the basis of a primitive nervous system. The firing rate of proteinoid microspheres can be used as an indicator of their ability to replicate the functions of a neuron, such as transmitting information. This provides evidence for the potential use of proteinoid microspheres as substrates for artificial neural networks. 

The strength of the linear model suggests that further research should focus on a deeper understanding of the underlying mechanism that leads to the correlation between molecular parameters and firing rate. Moreover, the linear model could be used to predict the firing rates of proteinoid microspheres for improved design of artificial neural networks. 

The linear model that best fits our data is represented by the following equation. Figure~\ref{bhjgljlkkjgkljbsdgsjbjbjk,,} shows the scatter plot of the firing rate versus the molecular weight and the peptide length, along with the regression plane of the model

\begin{figure}[!tbp]
\centering
\includegraphics[width=0.8\textwidth]{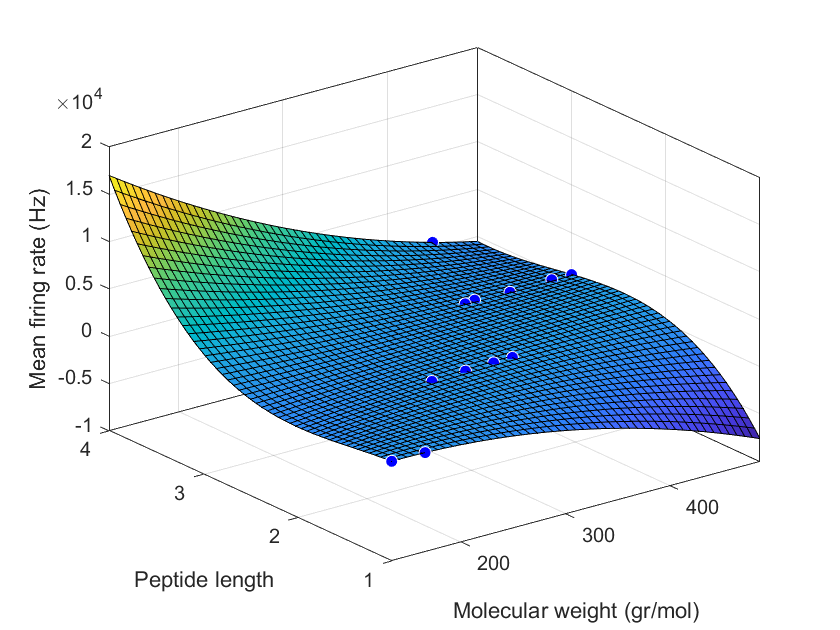}
\caption{For 12 distinct proteinoid microspheres, a QSAR model was used to predict the mean firing rates in Hz, peptide length, and molecular weight in g/mol.}
\label{bhjgljlkkjgkljbsdgsjbjbjk,,}
\end{figure}

\begin{align*}
f(x,y) &= p_{00} + p_{10}x + p_{01}y \\
&+ p_{20}x^2 + p_{11}xy + p_{02}y^2 \\
&+ p_{21}x^2y + p_{12}xy^2 + p_{03}y^3 \\
\text{Coefficients (with 95\% confidence bounds):} \\
p_{00} &= 2349 \; \left[-3560, 8258\right] \\
p_{10} &= -12.08 \; \left[-45.24, 21.07\right] \\
p_{01} &= -1770 \; \left[-1.172\times 10^4, 8182\right] \\
p_{20} &= -0.1149 \; \left[-0.6961, 0.4664\right] \\
p_{11} &= 48.49 \; \left[-128.7, 225.7\right] \\
p_{02} &= -2545 \; \left[-1.132\times 10^4, 6227\right] \\
p_{21} &= 0.04667 \; \left[-0.1628, 0.2561\right] \\
p_{12} &= -17.24 \; \left[-79.99, 45.5\right] \\
p_{03} &= 1151 \; \left[-2675, 4977\right]
\end{align*}

\begin{table}[h]
\centering
\caption{Mean firing rate, and predicted QSAR, of different proteinoid microsphere samples.}
\label{tab:qsar}
\resizebox{\textwidth}{!}{%
\begin{tabular}{lrrr}
\hline
Sample & Mean firing rate (Hz) & Predicted QSAR (Hz)  \\
\hline
L-Glu:L-Asp & 535.4877 & 536.0542   \\
L-Glu:L-Asp:L-Phe & 436.2721 & 492.8753   \\
L-Lys:L-Phe:L-Glu & 542.9443 & 563.6253 \\
L-Glu:L-Phe:L-His & 567.0562 & 521.7084 \\
L-Glu:L-Phe:PLLA & 498.2888 & 551.9483  \\
L-Lys:L-Phe:L-His:PLLA & 650.4798 & -901.3635& \\
L-Glu:L-Arg & 732.9516 & -2041.8  \\
L-Asp & 529.072 & 723.4966   \\
L-Phe:L-Lys & 768.2345 & -2619.1  \\
L-Glu:L-Asp:L-Pro & 617.3223 & 1345.4  \\
L-Phe & 491.5065 & 471.338   \\
L-Glu:L-Phe & 665.2995 & -1084.7   \\
\hline
\end{tabular}%
}
\end{table}

The present research indicates that proteinoid microspheres exhibiting higher mean firing rates, predicted QSAR, and \% deviations are more effective in transmitting signals than those with lower values. The observed phenomenon can be attributed to the increased capacity of the larger microspheres to accommodate a higher quantity of proteinoids, leading to a greater number of active neurons. Higher QSAR values suggest that microspheres are more likely to initiate a neuronal cascade, which is crucial for effective signal transmission. Mean firing rate is a crucial parameter for assessing the efficacy of a neuron in signal transmission, representing the average number of firings within a specified time frame. QSAR prediction refers to the anticipated capacity of a neuron to activate, derived from experimental data. The \% deviation represents the disparity between the anticipated outcome and the observed outcome of the experiment.
Proteinoid microspheres have potential as a substrate for developing artificial brains and unconventional computing devices due to their ability to generate and transmit electrical activity and react to external stimuli, as reported by certain sources~\cite{adamatzky2021towards}. They can form programmable networks through pores and tubes. This study of proteinoid oscillations provides insights into their molecular dynamics and intermolecular interactions.

\section{Discussion}

%The research findings suggest that proteinoid microspheres with higher mean firing rates, predicted QSAR, and % deviations are more effective in transmitting signals compared to those with lower values. This can be attributed to the larger microspheres' increased capacity to accommodate a greater quantity of proteinoids, resulting in a higher number of active neurons. Higher QSAR values indicate a higher likelihood of microspheres initiating a neuronal cascade, which is crucial for effective signal transmission. Mean firing rate is an important parameter for assessing a neuron's efficacy in signal transmission, representing the average number of firings within a specified time frame. QSAR prediction refers to the anticipated activation capacity of a neuron based on experimental data. % deviation represents the difference between the anticipated outcome and the observed outcome of the experiment.

%Proteinoid microspheres have the potential to be used as a substrate for developing artificial brains and unconventional computing devices. They possess the ability to generate and transmit electrical activity and respond to external stimuli, as reported by certain sources~\cite{adamatzky2021towards}. These microspheres can form programmable networks through pores and tubes. The study of proteinoid oscillations provides insights into their molecular dynamics and intermolecular interactions.

The findings of this paper shed light on the potential functions of proteinoids in neuronal circuitry, ranging from providing structure and format for electrical signals to acting as mediators in the transmission of physiological information. It is now possible to investigate communication in biological compounds through electrical oscillations and compare the results with those observed in more complex biological systems. The discussion section will delve deeper into the potential impact of this work on understanding the function of proteinoids in neuronal signaling and its implications for ongoing research into electrical communication in living organisms.

Recent research suggests a correlation between proteinoid oscillations and communication, similar to the correlation discovered by Adamatzky et al.~\cite{adamatzky2022language} in their investigation of oscillations in fungi. Communication between microspheres is crucial for the development and evolution of complex systems like unconventional computing and autonomous robotics. The nervous system of proteinoid microspheres and their analogues provide insights into their interactions.

Communication between microspheres primarily occurs through direct contact, allowing signal transmission through excitability. This involves the spheres coming into contact through surface tension or mechanical pressure (piezoelectricity). While this approach is reliable, chemical differences among the microspheres may hinder it. Electrical coupling is the most common method of communication between microspheres. It enables the transmission of binary information, such as digitally encoded data packets, using electrical signals. Capacitors, inductors, and transistors serve as connecting points between the microspheres. Magnetic coupling is another form of communication, where signals are exchanged through magnetic fields (ferroelectricity). This method utilizes an inductive current generated by one microsphere and received by the other. Optical coupling relies on the transmission of light for information exchange between microspheres~\cite{mougkogiannis2023light}. Light pulses are transmitted from one microsphere to another, allowing for relatively high data transfer rates. Communication among microspheres can be categorized into two distinct categories: information exchange and control. Information exchange involves the transmission and reception of data, while control refers to the transmission and reception of commands. Microspheres engage in information exchange, interaction, and resource sharing to facilitate the advancement of complex systems and procedures.

\begin{figure}[!tbp]
\centering
\includegraphics[width=1\textwidth]{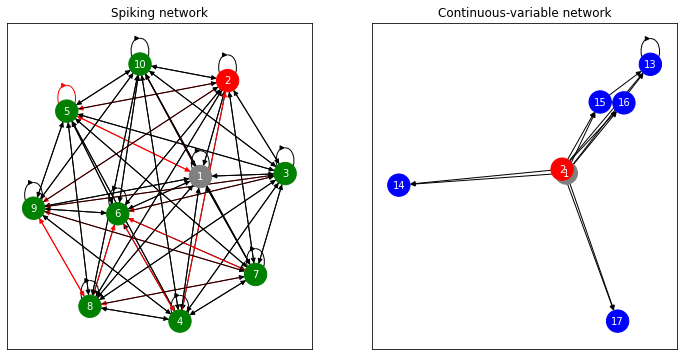}
\caption{Network architectures of proteinoid microspheres. a) Spiking network. A network of N recurrently connected leaky integrate-and-fire neurons (green circles) receives an input Fin(t) (grey circle) through synapses U, and generates an output Fout(t) (red circle) through synapses W.  b) Continuous-variable network. A network of $N_tilde$ recurrently connected ``rate" units (blue circles) receive inputs Fin(t) and Fout(t) through synapses $U_tilde$ and u, respectively~\cite{depasquale2016using}. }
\label{bhjglbjbjjbblhbjhbjk,,}
\end{figure}

Figure~\ref{bhjglbjbjjbblhbjhbjk,,} provides insights into potential interpretations and analogies of a proteinoid microsphere nervous system. It offers an understanding of microspheres' interactions and network organization similar to biological nervous systems. The figure presents two distinct architectures showcasing the potential functions of proteinoid microspheres. The first architecture depicts spiking networks composed of leaky integrate-and-fire neurons that receive external input force Fin(t) and produce output Fout(t) via synapses W. This architecture resembles the nervous system of advanced organisms, as the input and output signals exhibit similar behavior and generation patterns to those found in a typical nervous system. The second architecture utilizes continuous-variable networks to process an input from an external force Fin(t) and an internal output Fout(t) to produce the corresponding output. Continuous variables are employed instead of binary states of neurons, allowing for a wider range of interpretations and analogies of the nervous system. This network architecture provides a more realistic representation of the nervous system and its associated functions.

Proteinoid microspheres have the potential to function as protoneural networks, as shown in Figure~\ref{bhjglbjbjjbblhbjhbjk,,} with its two distinct architectures. Proteinoid microspheres serve as the fundamental units of the network, enabling basic communication and potential capacity for simple computations. As the network expands, it can develop intricate architectures that leverage the inherent connectivity of interconnected molecules, enabling significantly advanced functions and capabilities. This distinguishes proteinoid microsphere networks from conventional computing architectures that rely on external wiring for communication.

Proteinoid microspheres and biological neural networks share similarities in structure and function. Both consist of interconnected units capable of processing information and exhibiting adaptive behavior. Proteinoid microspheres are artificially synthesized structures made up of chemically bonded amino acids created in a laboratory setting. In contrast, neural networks are complex biological systems composed of individual neurons that work together in a coordinated manner. Proteinoid microspheres exhibit a less complex architecture compared to neural networks, with each node accountable for a single function, while neurons can process multiple inputs and perform various roles, such as transmitting signals between neurons or serving as synapses. Proteinoid microspheres have limited behavioral capabilities, primarily focused on simple tasks like self-repair and shape adaptation due to their inherent lack of complexity. In contrast, biological neural networks possess advanced capabilities such as memory formation, decision-making, and learning. While both proteinoid microspheres and biological neural networks can process information, the intricate and adaptable nature of biological neural networks sets them apart from proteinoid microspheres~\cite{fox1972molecular,miller1953production,miller1955production,dose1988origin}.

The results obtained from detecting spikes using differential pulse voltammetry (DPV) indicate that proteinoid microspheres could have had a significant impact on the emergence of life. Proteinoid microspheres possess the property of self-assembly, allowing for the aggregation of essential elements necessary for the genesis of protocells and the formation of intricate structures. Additionally, the microspheres have the ability to retain and convey data, a crucial prerequisite for the origin of biological existence. The collective capabilities of proteinoid microspheres may have facilitated the emergence and development of primitive cells during the initial phases of life.

%\begin{figure}[!tbp]
%\centering
%\includegraphics[width=1\textwidth]{figs/hipp4}
%\includegraphics[width=1\textwidth]{Reponse2stimulation.pdf}
%\caption{The Python code creates 1000 random numbers between 1 and 10000, plots them, and then computes and displays various firing pattern elements that may be seen in hippocampus neurons cultured in vitro. The components of the firing pattern are:Inter-spike interval (ISI): the time difference between two consecutive spikes. Post firing silence (PFS):is the data array's least value and the lowest spike amplitude.Slow depolarization wave (sDW):the data array's maximum value, or the largest spike amplitude. Slow after-hyperpolarization (sAHP): the variation between sDW and PFS that represents the amplitude of the slow hyperpolarizing phase following a spike. PFS is subtracted from sDW to calculate this.}
%\end{figure}

Proteinoid microspheres offer a fresh perspective for advancing our understanding of neural circuits. As researchers delve deeper into the system's adaptability, it is anticipated that proteinoids will unlock new insights in currently unexplored domains. These discoveries have the potential to pave the way for improved treatments for neurological disorders and advancements in medical technology and unconventional computing.

\section{Conclusion}

The findings of the study highlight the promising compatibility between differential pulse voltammetry and proteinoid nano-brains, opening up a new avenue for exploring these unique systems. The results suggest that utilizing differential pulse voltammetry as a tool can greatly contribute to understanding the functionality of proteinoid nano-brains, offering valuable insights into their behavior and potential applications. Further research in this area could unlock a deeper comprehension of these nano-brains and their potential role in the development of intelligent machines, potentially revolutionizing the field of artificial intelligence.

\section*{Acknowledgement}

The research was supported by EPSRC Grant EP/W010887/1 ``Computing with proteinoids''. Authors are grateful to David Paton for helping with SEM imaging and to Neil Phillips for helping with instruments.

%\bibliographystyle{plain}
%\bibliography{proteinoid_electrical_activity,mybibfile}
%\bibliography{proteinoid_electrical_activity,mybibfile}

\begin{thebibliography}{10}
\expandafter\ifx\csname url\endcsname\relax
  \def\url#1{\texttt{#1}}\fi
\expandafter\ifx\csname urlprefix\endcsname\relax\def\urlprefix{URL }\fi
\expandafter\ifx\csname href\endcsname\relax
  \def\href#1#2{#2} \def\path#1{#1}\fi

\bibitem{ji2019recent}
D.~Ji, T.~Li, W.~Hu, H.~Fuchs, Recent progress in aromatic polyimide
  dielectrics for organic electronic devices and circuits, Advanced Materials
  31~(15) (2019) 1806070.

\bibitem{fahlman2019interfaces}
M.~Fahlman, S.~Fabiano, V.~Gueskine, D.~Simon, M.~Berggren, X.~Crispin,
  Interfaces in organic electronics, Nature Reviews Materials 4~(10) (2019)
  627--650.

\bibitem{mao2019bio}
J.-Y. Mao, L.~Zhou, Y.~Ren, J.-Q. Yang, C.-L. Chang, H.-C. Lin, H.-H. Chou,
  S.-R. Zhang, Y.~Zhou, S.-T. Han, A bio-inspired electronic synapse using
  solution processable organic small molecule, Journal of Materials Chemistry C
  7~(6) (2019) 1491--1501.

\bibitem{matsui2019flexible}
H.~Matsui, Y.~Takeda, S.~Tokito, Flexible and printed organic transistors: From
  materials to integrated circuits, Organic Electronics 75 (2019) 105432.

\bibitem{feron2018organic}
K.~Feron, R.~Lim, C.~Sherwood, A.~Keynes, A.~Brichta, P.~C. Dastoor, Organic
  bioelectronics: materials and biocompatibility, International journal of
  molecular sciences 19~(8) (2018) 2382.

\bibitem{fox1992thermal}
S.~W. Fox, Thermal proteins in the first life and in the ``mind-body”
  problem, in: Evolution of Information Processing Systems, Springer, 1992, pp.
  203--228.

\bibitem{harada1958thermal}
K.~Harada, S.~W. Fox, The thermal condensation of glutamic acid and glycine to
  linear peptides1, Journal of the American Chemical Society 80~(11) (1958)
  2694--2697.

\bibitem{fox1995experimental}
S.~W. Fox, P.~R. Bahn, K.~Dose, K.~Harada, L.~Hsu, Y.~Ishima, J.~Jungck,
  J.~Kendrick, G.~Krampitz, J.~C. Lacey, et~al., Experimental retracement of
  the origins of a protocell, Journal of biological physics 20~(1-4) (1995)
  17--36.

\bibitem{rizzotti1998did}
M.~Rizzotti, M.~Crisma, F.~De~Luca, P.~Iobstraibizer, P.~Mazzei, Did the first
  cell emerge from a microsphere?, in: Exobiology: Matter, Energy, and
  Information in the Origin and Evolution of Life in the Universe, Springer,
  1998, pp. 199--202.

\bibitem{matsuno2012molecular}
K.~Matsuno, Molecular Evolution and Protobiology, Springer Science \& Business
  Media, 2012.

\bibitem{baluvska2021biomolecular}
F.~Balu{\v{s}}ka, W.~B. Miller~Jr, A.~S. Reber, Biomolecular basis of cellular
  consciousness via subcellular nanobrains, International Journal of Molecular
  Sciences 22~(5) (2021) 2545.

\bibitem{callaway2021deepmind}
E.~Callaway, et~al., Deepmind’s ai predicts structures for a vast trove of
  proteins, Nature 595~(7869) (2021) 635--635.

\bibitem{2018Artificial}
Artificial neurons compute faster than the human brain,
  https://www.nature.com/articles/d41586-018-01290-0 (jan 26 2018).

\bibitem{zhang2019artificial}
Q.~Zhang, H.~Yu, M.~Barbiero, B.~Wang, M.~Gu, Artificial neural networks
  enabled by nanophotonics, Light: Science \& Applications 8~(1) (2019) 42.

\bibitem{nwadiugwu2020neural}
M.~C. Nwadiugwu, Neural networks, artificial intelligence and the computational
  brain, arXiv preprint arXiv:2101.08635 (2020).

\bibitem{fox1982updated}
S.~W. Fox, T.~Nakashima, A.~Przybylski, R.~M. Syren, The updated experimental
  proteinoid model, International Journal of Quantum Chemistry 22~(S9) (1982)
  195--204.

\bibitem{goi2020perspective}
E.~Goi, Q.~Zhang, X.~Chen, H.~Luan, M.~Gu, Perspective on photonic memristive
  neuromorphic computing, PhotoniX 1 (2020) 1--26.

\bibitem{fiers2013nanophotonic}
M.~A.~A. Fiers, T.~Van~Vaerenbergh, F.~Wyffels, D.~Verstraeten, B.~Schrauwen,
  J.~Dambre, P.~Bienstman, Nanophotonic reservoir computing with photonic
  crystal cavities to generate periodic patterns, IEEE transactions on neural
  networks and learning systems 25~(2) (2013) 344--355.

\bibitem{kwon2016neuron}
E.~J. Kwon, M.~Skalak, R.~Lo~Bu, S.~N. Bhatia, Neuron-targeted nanoparticle for
  sirna delivery to traumatic brain injuries, ACS nano 10~(8) (2016)
  7926--7933.

\bibitem{ramachandran2017theranostic}
R.~Ramachandran, V.~R. Junnuthula, G.~S. Gowd, A.~Ashokan, J.~Thomas,
  R.~Peethambaran, A.~Thomas, A.~K.~K. Unni, D.~Panikar, S.~V. Nair, et~al.,
  Theranostic 3-dimensional nano brain-implant for prolonged and localized
  treatment of recurrent glioma, Scientific reports 7~(1) (2017) 43271.

\bibitem{mougkogiannis2023low}
P.~Mougkogiannis, A.~Adamatzky, Low frequency electrical waves in ensembles of
  proteinoid microspheres, Scintific Reports 13 (2023) 1992.

\bibitem{syed2023atomically}
G.~S. Syed, Y.~Zhou, J.~Warner, H.~Bhaskaran, Atomically thin optomemristive
  feedback neurons, Nature Nanotechnology (2023) 1--8.

\bibitem{wan2020artificial}
C.~Wan, P.~Cai, X.~Guo, M.~Wang, N.~Matsuhisa, L.~Yang, Z.~Lv, Y.~Luo, X.~J.
  Loh, X.~Chen, An artificial sensory neuron with visual-haptic fusion, Nature
  communications 11~(1) (2020) 4602.

\bibitem{yegnanarayana1994artificial}
B.~Yegnanarayana, Artificial neural networks for pattern recognition, Sadhana
  19 (1994) 189--238.

\bibitem{mougkogiannis2023transfer}
P.~Mougkogiannis, N.~Phillips, A.~Adamatzky, Transfer functions of proteinoid
  microspheres, arXiv preprint arXiv:2302.05255 (2023).

\bibitem{rabbani2017dynamic}
A.~Rabbani, S.~Salehi, Dynamic modeling of the formation damage and mud cake
  deposition using filtration theories coupled with sem image processing,
  Journal of Natural Gas Science and Engineering 42 (2017) 157--168.

\bibitem{kon2000information}
M.~A. Kon, L.~Plaskota, Information complexity of neural networks, Neural
  Networks 13~(3) (2000) 365--375.

\bibitem{HedgesElectrical}
V.~Hedges, Electrical {Activity} in {Neurons} -- {Introduction} to
  {Neuroscience},
  https://openbooks.lib.msu.edu/introneuroscience1/chapter/postsynaptic-potentials/.

\bibitem{MIT2018Seeing}
A.~T. MIT News~Office, Seeing the brain's electrical activity,
  https://news.mit.edu/2018/seeing-brains-electrical-activity-0226 (Feb 26
  2018).

\bibitem{xu2018collective}
Y.~Xu, Y.~Jia, J.~Ma, T.~Hayat, A.~Alsaedi, Collective responses in electrical
  activities of neurons under field coupling, Scientific reports 8~(1) (2018)
  1349.

\bibitem{pinotsis2023cytoelectric}
D.~A. Pinotsis, G.~Fridman, E.~K. Miller, Cytoelectric coupling: Electric
  fields sculpt neural activity and “tune” the brain’s infrastructure,
  Progress in Neurobiology (2023) 102465.

\bibitem{ouldali2020electrical}
H.~Ouldali, K.~Sarthak, T.~Ensslen, F.~Piguet, P.~Manivet, J.~Pelta, J.~C.
  Behrends, A.~Aksimentiev, A.~Oukhaled, Electrical recognition of the twenty
  proteinogenic amino acids using an aerolysin nanopore, Nature biotechnology
  38~(2) (2020) 176--181.

\bibitem{rosenberg1962electrical}
B.~Rosenberg, Electrical conductivity of proteins, Nature 193~(4813) (1962)
  364--365.

\bibitem{gurth2020synaptic}
C.-M. G{\"u}rth, T.~M. Dankovich, S.~O. Rizzoli, E.~D’Este, Synaptic activity
  and strength are reflected by changes in the post-synaptic secretory pathway,
  Scientific Reports 10~(1) (2020) 1--13.

\bibitem{adamatzky2021towards}
A.~Adamatzky, Towards proteinoid computers. hypothesis paper, Biosystems 208
  (2021) 104480.

\bibitem{adamatzky2022language}
A.~Adamatzky, Language of fungi derived from their electrical spiking activity,
  Royal Society Open Science 9~(4) (2022) 211926.

\bibitem{mougkogiannis2023light}
P.~Mougkogiannis, A.~Adamatzky, Light induced spiking of proteinoids, arXiv
  preprint arXiv:2303.17563 (2023).

\bibitem{depasquale2016using}
B.~DePasquale, M.~M. Churchland, L.~Abbott, Using firing-rate dynamics to train
  recurrent networks of spiking model neurons, arXiv preprint arXiv:1601.07620
  (2016).

\bibitem{fox1972molecular}
S.~W. Fox, K.~Dose, Molecular evolution and the origin of life (1972).

\bibitem{miller1953production}
S.~L. Miller, A production of amino acids under possible primitive earth
  conditions, Science 117~(3046) (1953) 528--529.

\bibitem{miller1955production}
S.~L. Miller, Production of some organic compounds under possible primitive
  earth conditions, J. Am. Chem. Soc 77~(9) (1955) 2351--2361.

\bibitem{dose1988origin}
K.~Dose, The origin of life: More questions than answers, Interdisciplinary
  Science Reviews 13~(4) (1988) 348--356.

\end{thebibliography}

\end{document}

\begin{frontmatter}

\title{Supplementary materials. \\Spiking frequency modulation of proteinoids with light}

%% Authors and affiliations

\author{Panagiotis Mougkogiannis*}
\author{Andrew Adamatzky}
\address{Unconventional Computing Laboratory, UWE, Bristol, UK}

\end{frontmatter}

\begin{figure}[H]
\centering
\includegraphics[width=1\textwidth]{figs/frequency of periods glu_phe.png}
\caption{The L-Glu:L-Phe proteinoid oscillations have a period of 128 seconds and an average interval between maxima of 1650.9 seconds, as shown by the a) electrical oscillations, b) histogram of periods, c) histogram of amplitudes, and d) Fourier transform of data. Illumination  with cold white light from 37.2  to 186.6 klux (30 min ON - 30 min OFF).
}
\label{cavfsdvsd}
\end{figure}

\begin{figure}[H]
\centering
\includegraphics[width=1\textwidth]{figs/frequency of periods phe lys.png}
\caption{The L-Phe:L-Lys proteinoid oscillations have a period of 42.6667 seconds and an average interval between maxima of 3424.7 seconds, as shown by the a) electrical oscillations, b) histogram of periods, c) histogram of amplitudes, and d) Fourier transform of data. Illumination  with cold white light from 37.2 to 186.6 klux (30 min ON - 30 min OFF).
}
\label{ppsoiunpsssazx999}
\end{figure}

\begin{figure}[H]
\centering
\includegraphics[width=1\textwidth]{figs/frequency of periods Phe.png}
\caption{The L-Phe proteinoid oscillations have a period of 128 seconds and an average interval between maxima of 850.44 seconds, as shown by the a) electrical oscillations, b) histogram of periods, c) histogram of amplitudes, and d) Fourier transform of data.  Illumination  with cold white light. Illumination  with cold white light from 37.2  to 186.6 klux (30 min ON - 30 min OFF).
}
\label{ppsoiunpsssazx999}
\end{figure}

\begin{figure}[!tbp]
\centering
\includegraphics[width=1\textwidth]{figs/phe-lys-black white.png}
\caption{The L-Phe:L-Lys proteinoid oscillations have a period of 170.6667 seconds and an average interval between maxima of 4047.3 seconds, as shown by the a) electrical oscillations, b) histogram of periods, c) histogram of amplitudes, and d) Fourier transform of data. Illumination  with cold white light (37.2 klux) and black light (695.8 lux)  (30 min ON).
}
\label{ppsoiunpsssazx999}
\end{figure}

\begin{figure}[!tbp]
\centering
\includegraphics[width=1\textwidth]{figs/glu-phe-his black white.png}
\caption{The L-Glu:L-Phe:L-His proteinoid oscillations have a period of 5.5054 seconds and an average interval between maxima of 4121.7 seconds, as shown by the a) electrical oscillations, b) histogram of periods, c) histogram of amplitudes, and d) Fourier transform of data. Illumination  with cold white light (37.2 klux) and black light (695.8 lux - 30 min ON).
}
\label{vgbjbjk,n}
\end{figure}

\begin{figure}[!tbp]
\centering
\includegraphics[width=1\textwidth]{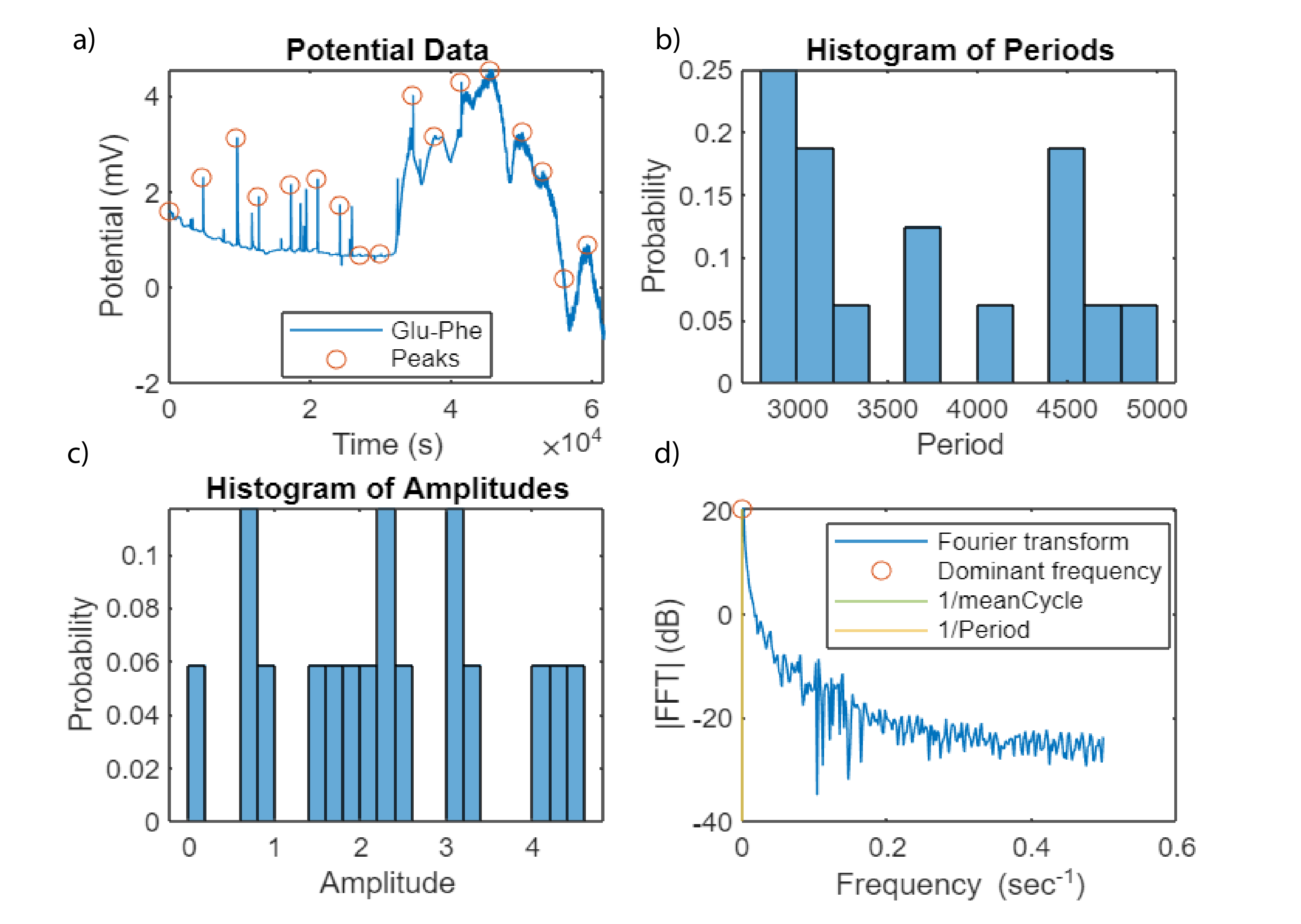}
\caption{The L-Glu:L-Phe proteinoid oscillations have a period of 512 seconds and an average interval between maxima of 3692.9 seconds, as shown by the a) electrical oscillations, b) histogram of periods, c) histogram of amplitudes, and d) Fourier transform of data. Illumination  with cold white light (37.2 klux) and black light (695.8 lux-30 min ON).
}
\label{gffghfvkhj}
\end{figure}

\begin{figure}[!tbp]
\centering
\includegraphics[width=1\textwidth]{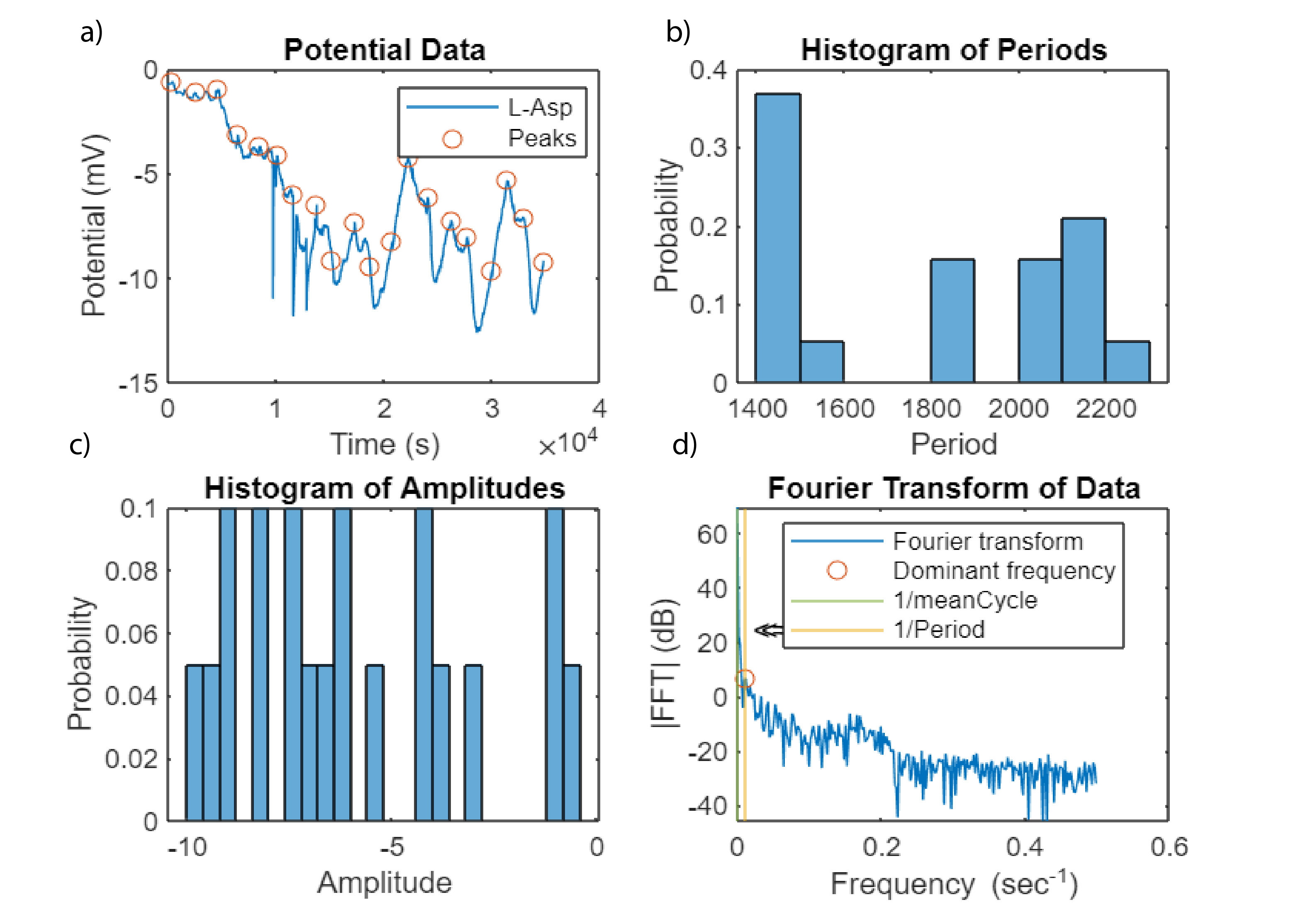}
\caption{The L-Asp proteinoid oscillations have a period of 85.3333 seconds and an average interval between maxima of 1807.3 seconds, as shown by the a) electrical oscillations, b) histogram of periods, c) histogram of amplitudes, and d) Fourier transform of data. Illumination  with cold white (37.2 klux) light and black light (695.8 lux-30 min ON).
}
\label{fdsgsgas}
\end{figure}

\begin{figure}[!tbp]
\centering
\includegraphics[width=1\textwidth]{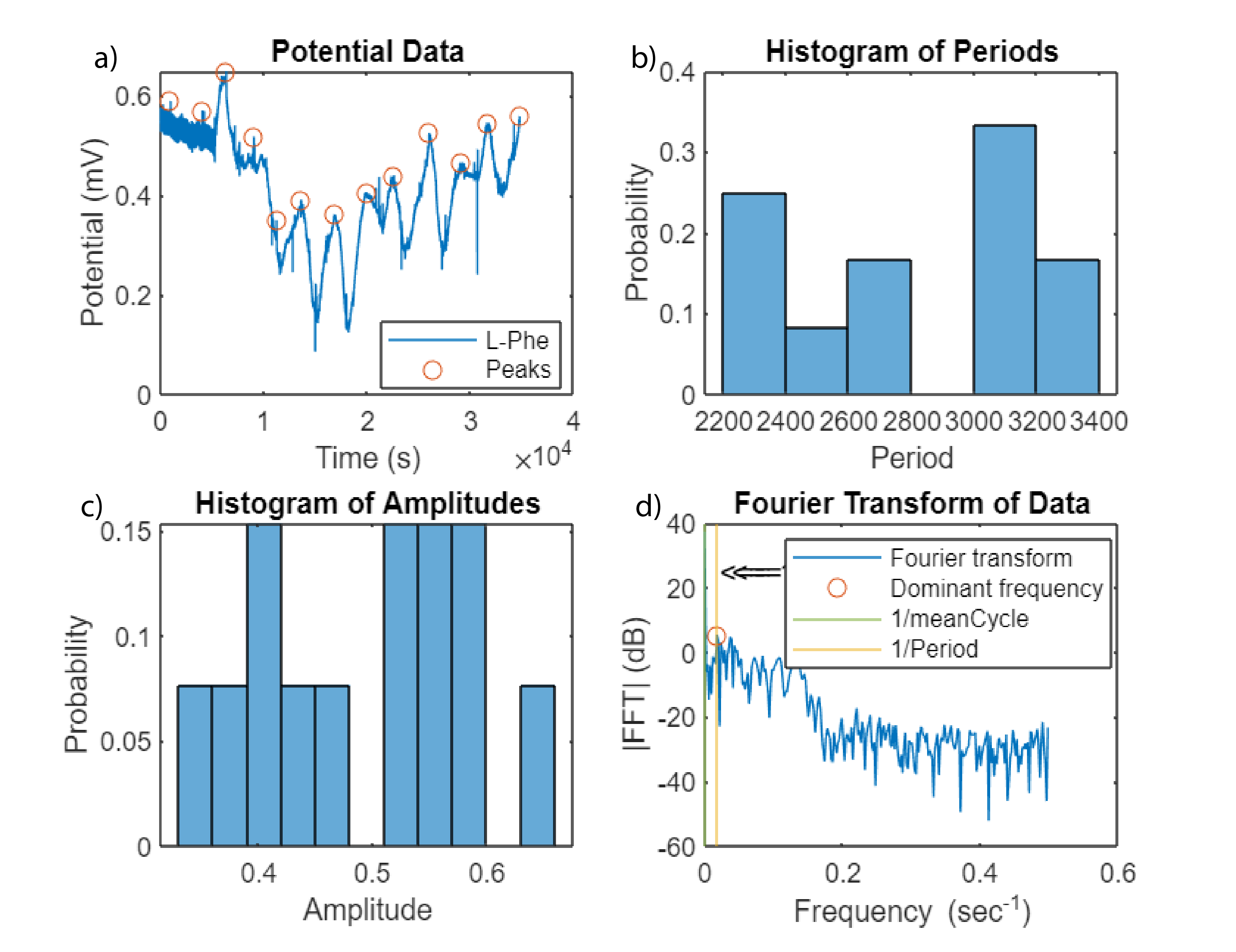}
\caption{The L-Phe proteinoid oscillations have a period of 56.8889 seconds and an average interval between maxima of 2813.9 seconds, as shown by the a) electrical oscillations, b) histogram of periods, c) histogram of amplitudes, and d) Fourier transform of data. Illumination  with cold white light (37.2 klux) and black light (695.8 lux-30 min ON).
}
\label{asdgfasgghh}
\end{figure}

\begin{figure}[!tbp]
\centering
\includegraphics[width=1\textwidth]{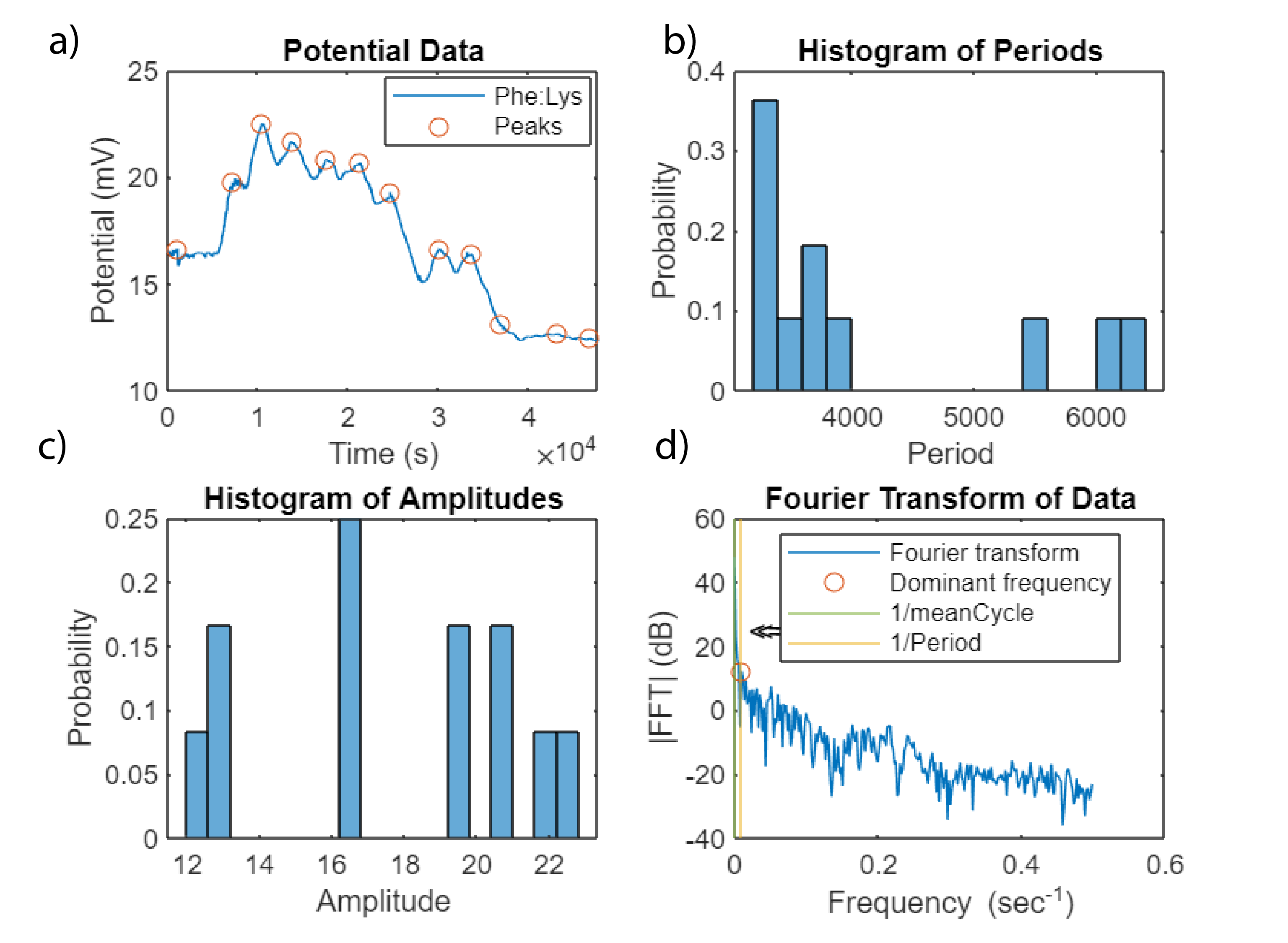}
\caption{The L-Phe:L-Lys proteinoid oscillations have a period of 102.4 seconds and an average interval between maxima of 4154.73  seconds, as shown by the a) electrical oscillations, b) histogram of periods, c) histogram of amplitudes, and d) Fourier transform of data. Illumination  with cold white light (68 klux) and black light (30 min ON).
}
\label{dsgsgsgsgsggsdgs}
\end{figure}

\begin{figure}[!tbp]
\centering
\includegraphics[width=1\textwidth]{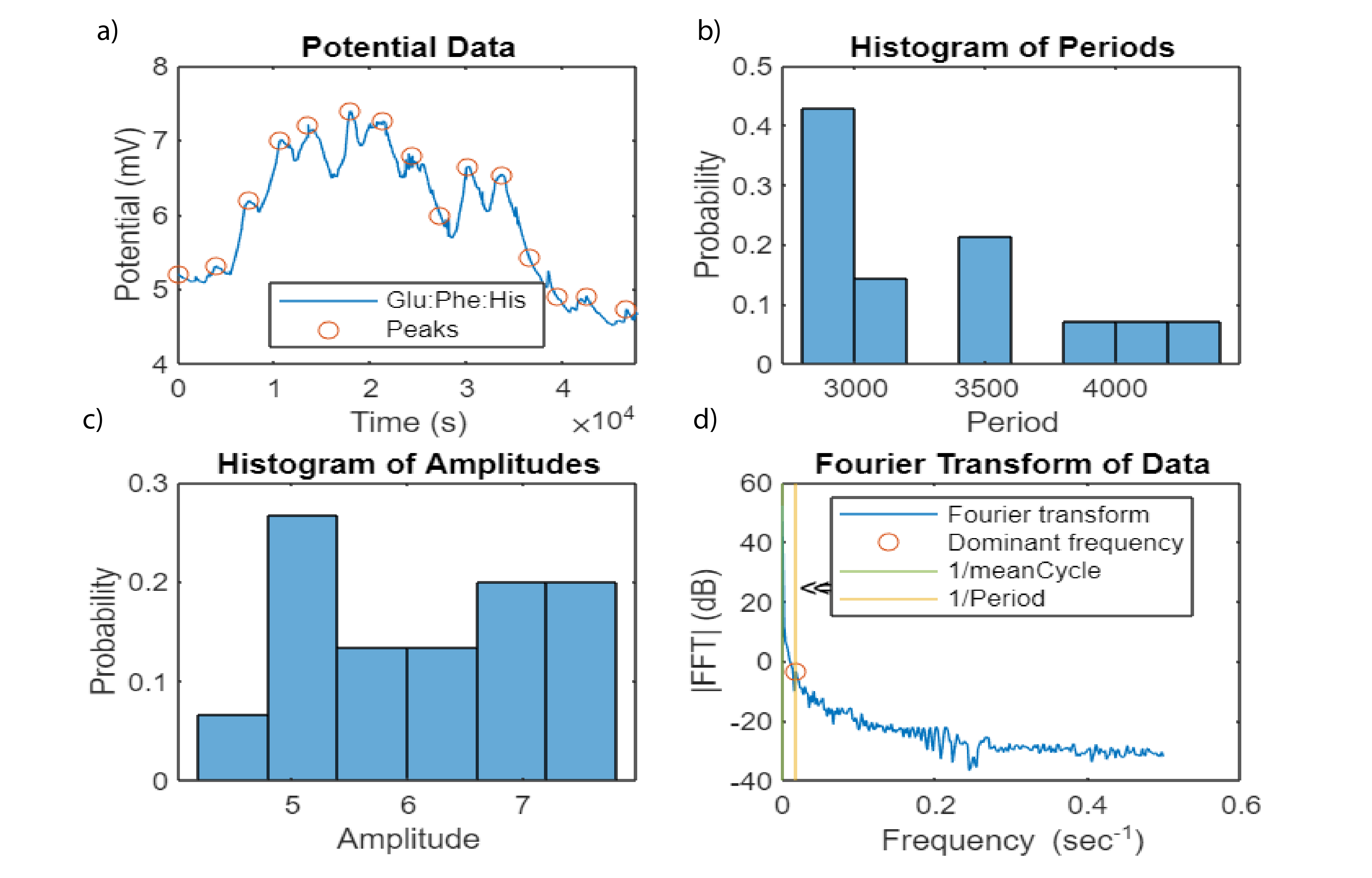}
\caption{The L-Glu:L-Phe:L-His proteinoid oscillations have a period of 56.8889 seconds and an average interval between maxima of 3328.6 seconds, as shown by the a) electrical oscillations, b) histogram of periods, c) histogram of amplitudes, and d) Fourier transform of data. Illumination  with cold white light (68 klux) and black light (30 min ON).
}
\label{fdsafsdagsagvasg}
\end{figure}

\begin{figure}[!tbp]
\centering
\includegraphics[width=1\textwidth]{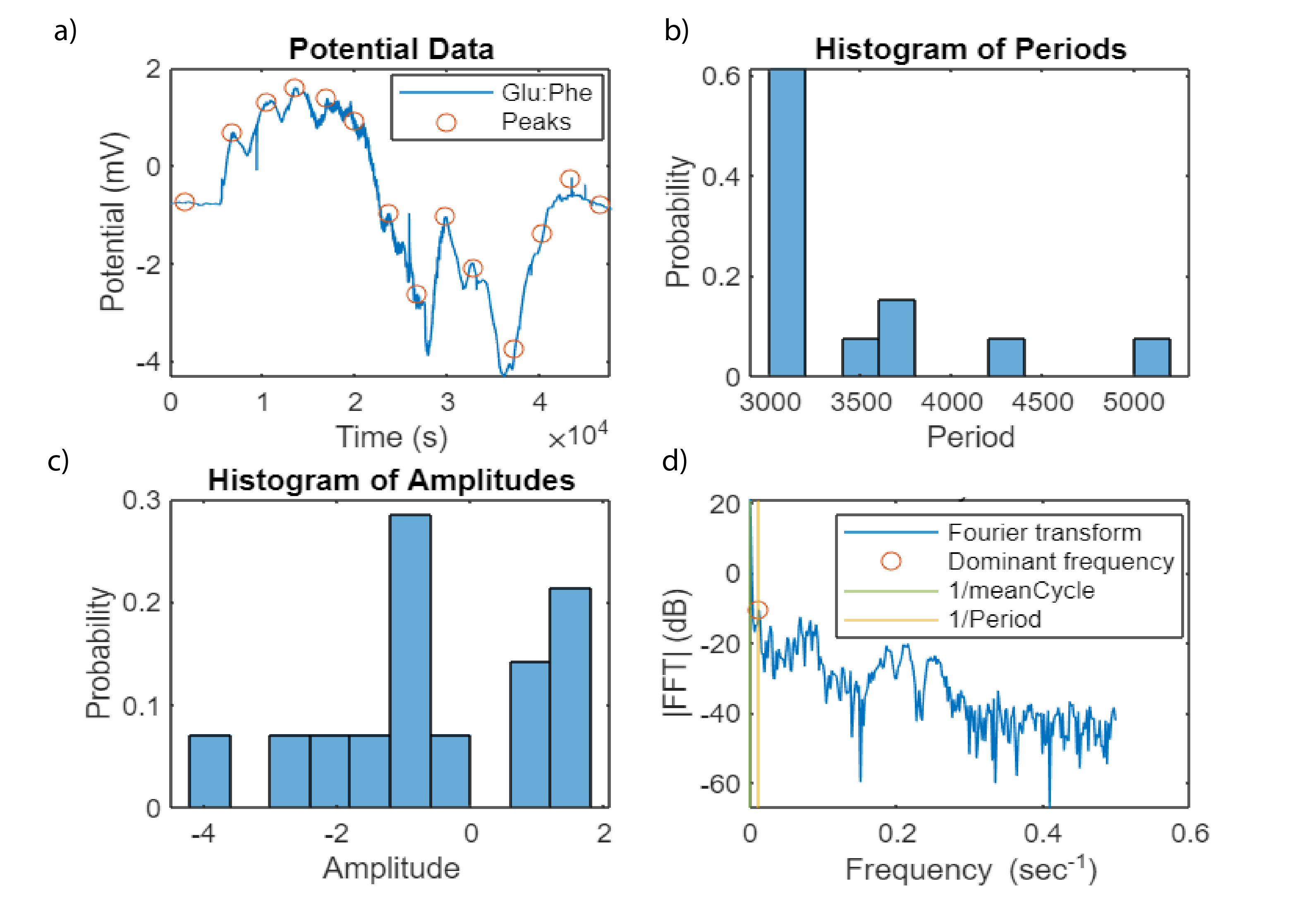}
\caption{The L-Glu:L-Phe proteinoid oscillations have a period of 636.41 seconds and an average interval between maxima of 3450.08  seconds, as shown by the a) electrical oscillations, b) histogram of periods, c) histogram of amplitudes, and d) Fourier transform of data. Illumination  with cold white light (68 klux) and black light (30 min ON).
}
\label{iohoghsioghsanvj}
\end{figure}

\begin{figure}[!tbp]
\centering
\includegraphics[width=1\textwidth]{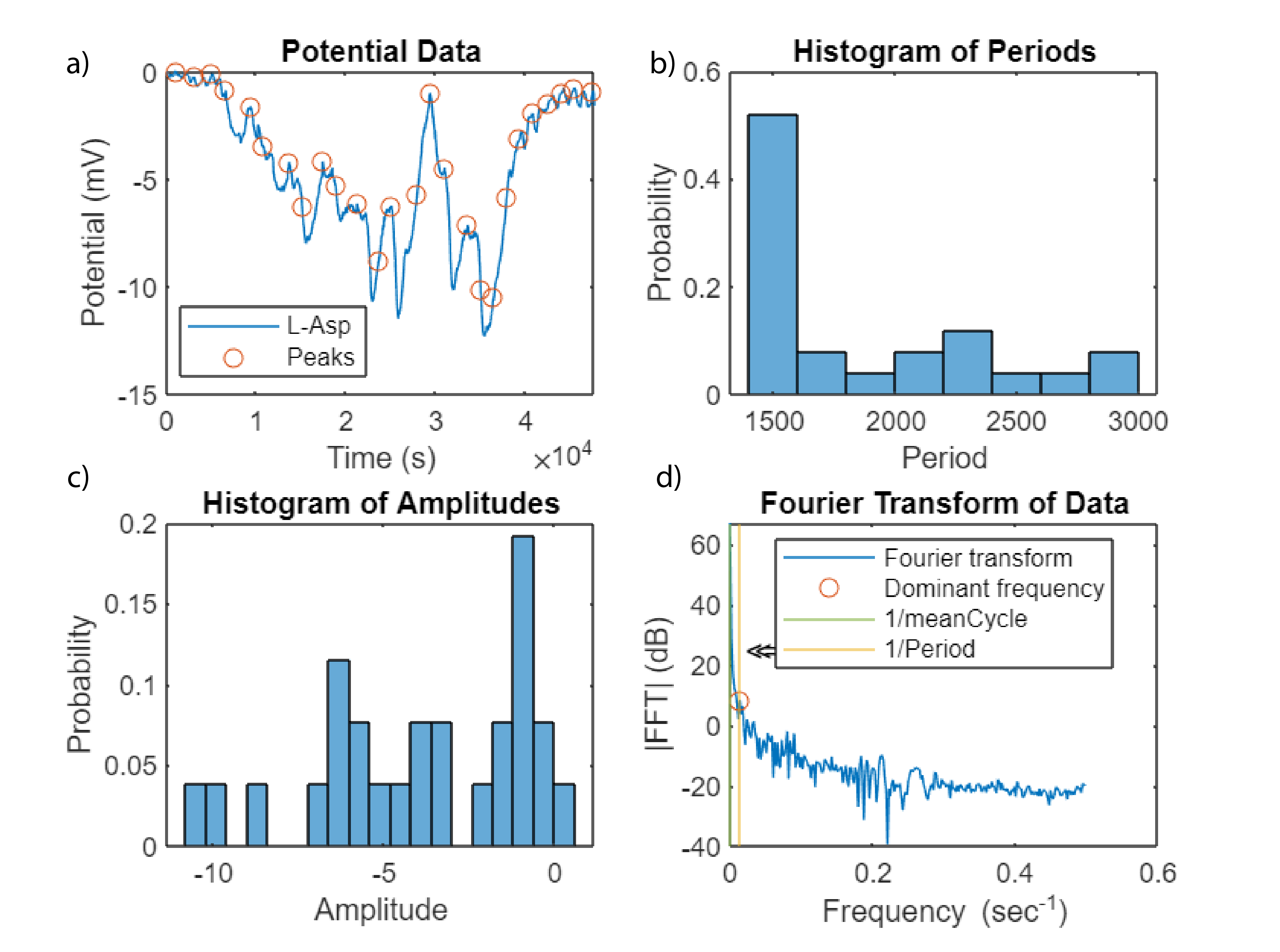}
\caption{The L-Asp proteinoid oscillations have a period of  73.1429 seconds and an average interval between maxima of  1541.00 seconds, as shown by the a) electrical oscillations, b) histogram of periods, c) histogram of amplitudes, and d) Fourier transform of data. Illumination  with cold white light (68 klux) and black light (30 min ON).
}
\label{sdghsdhdjsgfjsfkjk}
\end{figure}

\begin{figure}[!tbp]
\centering
\includegraphics[width=1\textwidth]{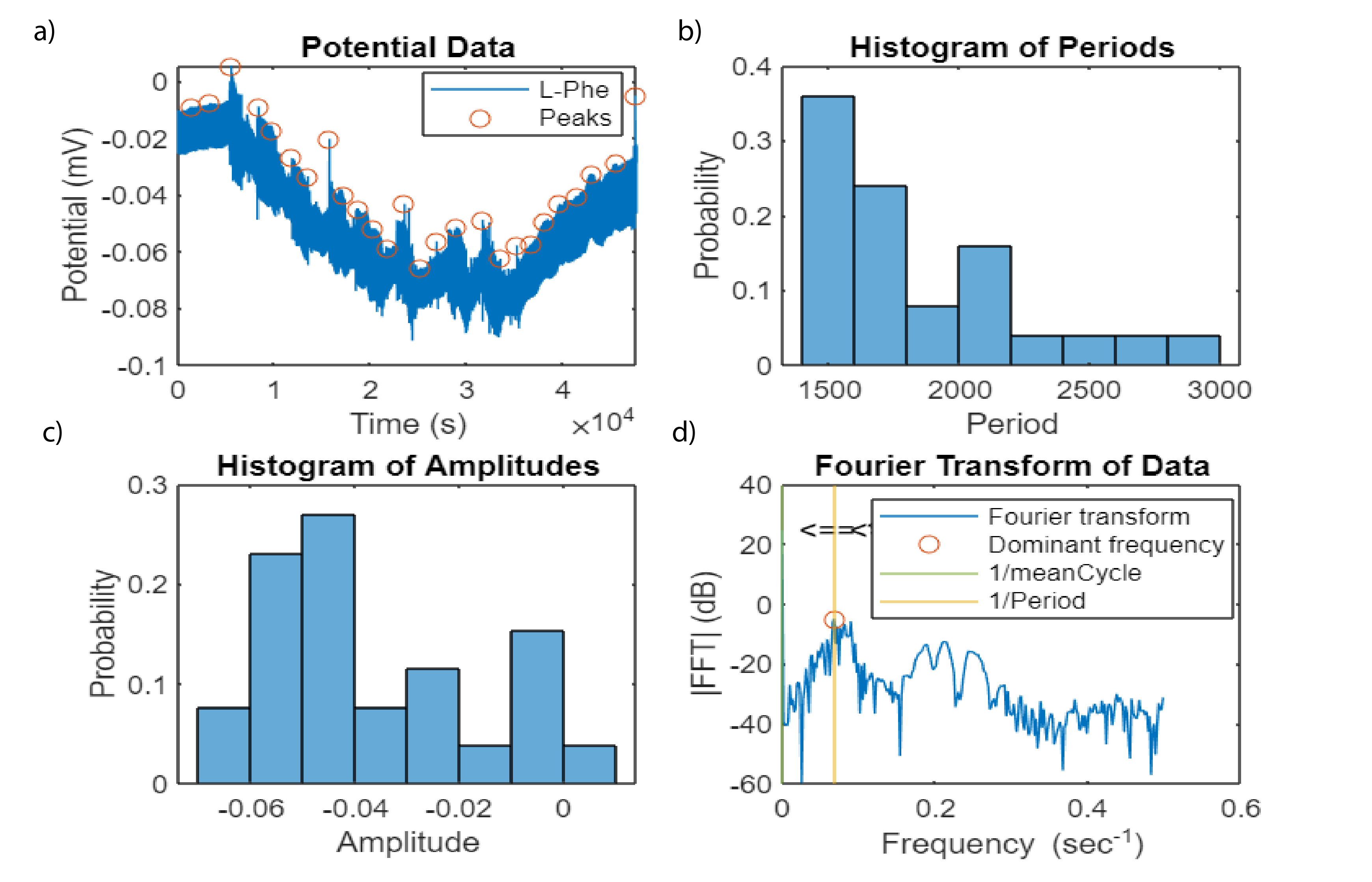}
\caption{The L-Phe proteinoid oscillations have a period of 14.6286 seconds and an average interval between maxima of  1844.12 seconds, as shown by the a) electrical oscillations, b) histogram of periods, c) histogram of amplitudes, and d) Fourier transform of data. Illumination  with cold white light (68 klux) and black light (30 min ON).
}
\label{sdghsdhdjsgfjsfkjk}
\end{figure}

\begin{figure}[!tbp]
\centering
\includegraphics[width=1\textwidth]{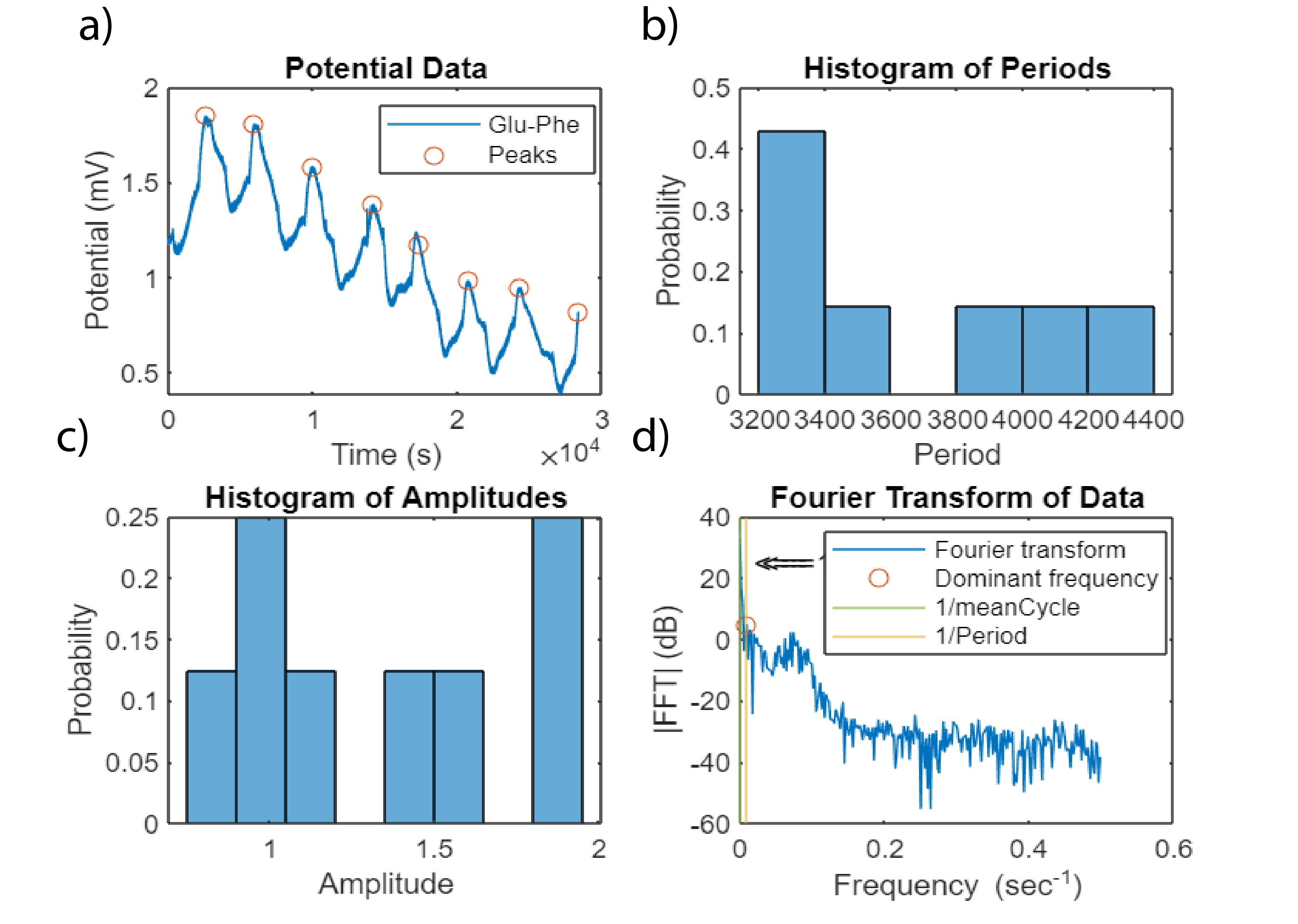}
\caption{The L:Glu:L-Phe proteinoid oscillations have a dominant period of 102.4 seconds and an average interval between maxima of 3.6766E+3 seconds, as shown by the a) electrical oscillations, b) histogram of periods, c) histogram of amplitudes, and d) Fourier transform of data.  Illumination  with cold white light (186.6 klux). Illumination  with cold white light (30 min ON - 30 min OFF)
}
\label{jfjsdnvlasv}
\end{figure}
\begin{figure}[!tbp]
\centering
\includegraphics[width=1\textwidth]{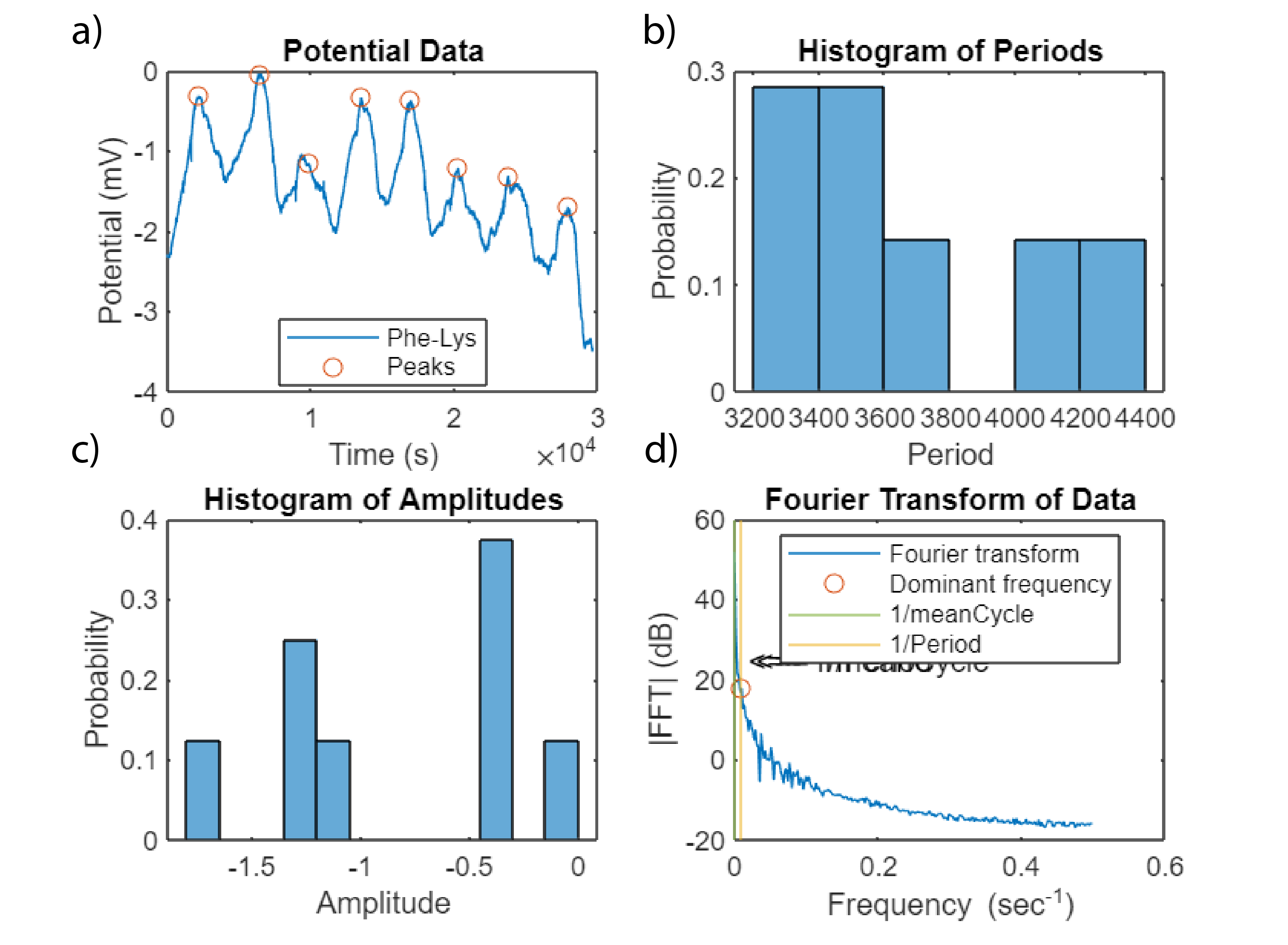}
\caption{The L:Phe:L-Lys proteinoid oscillations have a dominant period of 102.4 seconds and an average interval between maxima of 3.6761E+3 seconds, as shown by the a) electrical oscillations, b) histogram of periods, c) histogram of amplitudes, and d) Fourier transform of data.  Illumination  with cold white light (160.8 klux). Illumination  with cold white light (30 min ON - 30 min OFF)
}
\label{mdmvksmavkads}
\end{figure}

\begin{figure}[!tbp]
\centering
\includegraphics[width=1\textwidth]{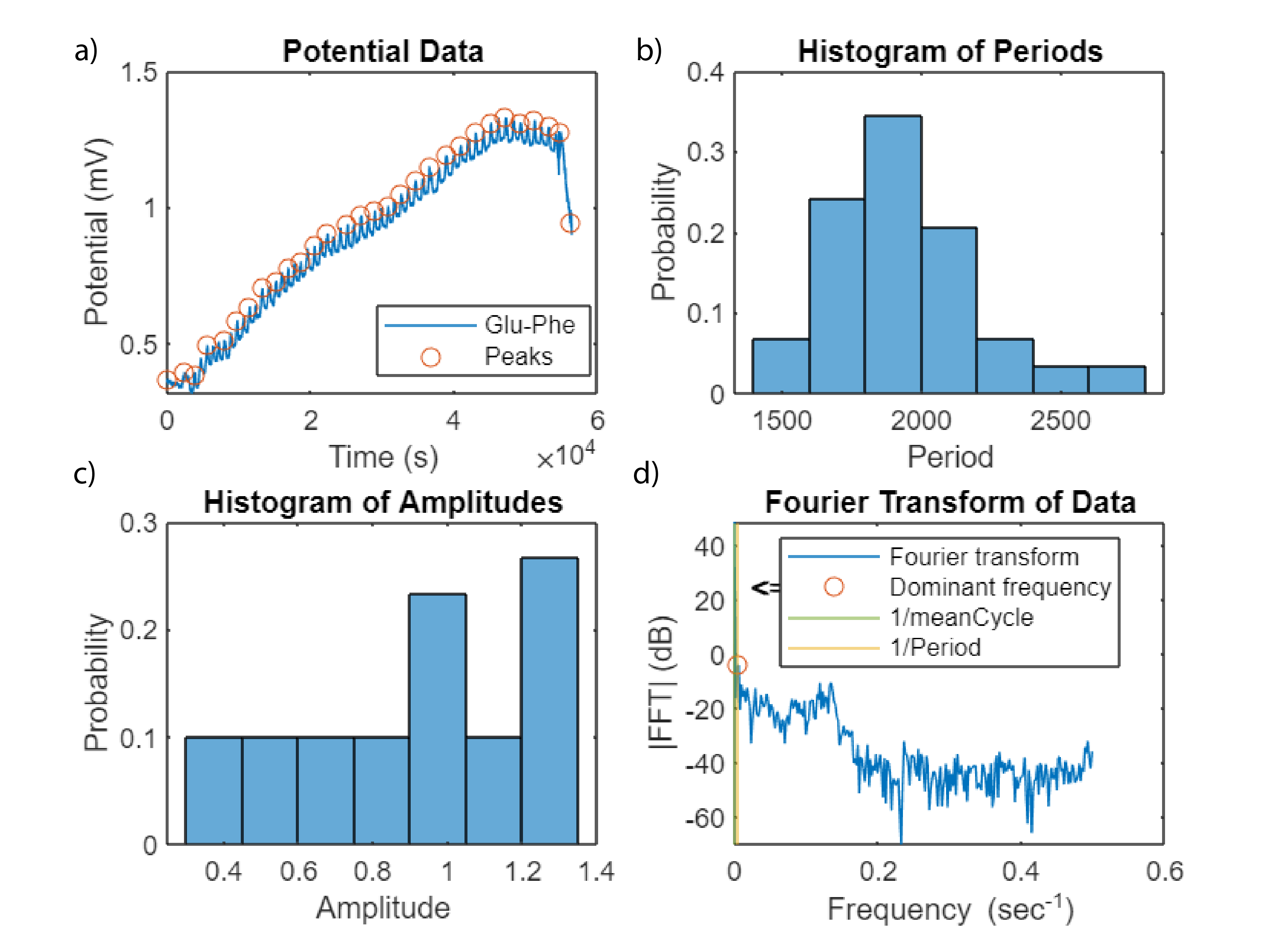}
\caption{The L:Glu:L-Phe proteinoid oscillations have a dominant period of 283.12 seconds and an average interval between maxima of 1.9393E+3 seconds, as shown by the a) electrical oscillations, b) histogram of periods, c) histogram of amplitudes, and d) Fourier transform of data.  Illumination  with black light (695.8 lux). 
}
\label{mdmvksmavkads}
\end{figure}

\begin{table}[!tbp]
\centering
\caption{The Protenoids L-Glu:L-Phe:L-His, L-Glu:L-Phe:L-Lys, and L-Phe were illuminated with cold black light (695.8 lux)  for 8 hours, and the peak voltage and period were recorded.}
\begin{tabular}{|c|c|c|c|}
	\hline
	Proteinoid  & Light & Peaks & Periods      \\
       &  Condition  &  Mean  &  Mean     \\
    &  & (mV) & (sec)      \\
	\hline\hline
	
	L-Glu:L-Phe:L-His& Constant & 7.54$\pm$2.08  &  941.98$\pm$145.81   \\
 L-Glu:L-Phe & Constant & 0.93$\pm$0.31 & 1929.28$\pm$283.12    \\
 L-Phe:L-Lys & Constant & 5.66$\pm$3.95& 12074$\pm$1123.94      \\
 L-Phe & Constant & $-$0.11$\pm$0.0  & 6272.88$\pm$16.84.37      \\
 L-Asp & Constant & $-$13.55$\pm$5.06  &   5036.18$\pm$1087.01     \\
	\hline
	\end{tabular}
\label{lksfagdsgsdgag}
\end{table}

\begin{table}[!tbp]
\centering
\caption{The Protenoids L-Glu:L-Phe:L-His, L-Glu:L-Phe:L-Lys, and L-Phe were illuminated with cold black light (695.8 lux)  and cold white light (37.2 klux) for 30 minute on/30 minute off cycles, and the peak voltage and period were recorded.}
\begin{tabular}{|c|c|c|c|}
	\hline
	Proteinoid  & Light & Peaks & Periods      \\
       &  Condition  &  Mean  &  Mean     \\
    &  & (mV) & (sec)      \\
	\hline\hline
	
	L-Glu:L-Phe:L-His& Periodic & 6.85$\pm$0.67  &  4068.57$\pm$1139.96    \\
 L-Glu:L-Phe & Periodic & 2.56$\pm$1.16 &  4220.50$\pm$965.87  \\
 L-Phe:L-Lys & Periodic & 22.28$\pm$2.74 &  3541.38$\pm$861.05    \\
 L-Phe & Periodic & 0.45$\pm$0.12  &    2251.13$\pm$623.43   \\
 L-Asp & Periodic & $-$6.12$\pm$3.12  & 1807.84$\pm$198.32      \\
	\hline
	\end{tabular}
\label{gfdjfhk}
\end{table}

\begin{table}[!tbp]
\centering
\caption{The Protenoids L-Glu:L-Phe:L-His, L-Glu:L-Phe:L-Lys, and L-Phe were illuminated with cold black light (695.8 lux) and cold white light (55.9 klux) for 30 minute on/30 minute off cycles, and the peak voltage and period were recorded.}
\begin{tabular}{|c|c|c|c|}
	\hline
	Proteinoid  & Light & Peaks & Periods      \\
       &  Condition  &  Mean  &  Mean     \\
    &  & (mV) & (sec)      \\
	\hline\hline
	
	L-Glu:L-Phe:L-His& Periodic &  6.10$\pm$0.95 & 3328.64$\pm$514.77      \\
 L-Glu:L-Phe & Periodic & $-$0.55$\pm$1.61  &  3450.08$\pm$636.41  \\
 L-Phe:L-Lys & Periodic & 17.73$\pm$3.59 & 4154.73$\pm$1176.58      \\
 L-Phe & Periodic &  $-$0.04$\pm$0.02 & 1844.12$\pm$404.17       \\
 L-Asp & Periodic & $-$3.87$\pm$3.14  &   1855.72$\pm$511.80     \\
	\hline
	\end{tabular}
\label{jksdbjkghjasdbjlgs}
\end{table}

\begin{table}[!tbp]
\centering
\caption{The Protenoids L-Glu:L-Phe:L-His, L-Glu:L-Phe:L-Lys, and L-Phe were illuminated with cold black light (695.8 lux) and cold white light (101.8 klux) for 30 minute on/30 minute off cycles, and the peak voltage and period were recorded.}
\begin{tabular}{|c|c|c|c|}
	\hline
	Proteinoid  & Light & Peaks & Periods      \\
       &  Condition  &  Mean  &  Mean     \\
    &  & (mV) & (sec)      \\
	\hline\hline
	
	L-Glu:L-Phe:L-His& Periodic &  7.68$\pm$1.00 & 3359$\pm$604.21      \\
 L-Glu:L-Phe & Periodic &  0.50$\pm$0.09 & 3412.75$\pm$367.04    \\
 L-Phe:L-Lys & Periodic &  13.09 $\pm$1.46 &  3605$\pm$628.72     \\
 L-Phe & Periodic & 0.08$\pm$0.16  & 4772.33$\pm$1715.47      \\
 L-Asp & Periodic & $-$1.79$\pm$2.93  & 1401.76$\pm$307.57       \\
	\hline
	\end{tabular}
\label{fghfgfhkj}
\end{table}

\begin{table}[!tbp]
\centering
\caption{The Protenoids L-Glu:L-Phe:L-His, L-Glu:L-Phe,L-Phe:L-Lys, L-Asp, and L-Phe were illuminated with cold black light (695.8 lux) and cold white light (186.6 klux) for 30 minute on/30 minute off cycles, and the peak voltage and period were recorded.}
\begin{tabular}{|c|c|c|c|}
	\hline
	Proteinoid  & Light & Peaks & Periods      \\
       &  Condition  &  Mean  &  Mean     \\
    &  & (mV) & (sec)      \\
	\hline\hline
	
	L-Glu:L-Phe:L-His& Periodic & $-$0.88$\pm$2.42 & 3382.25$\pm$704.92      \\
 L-Glu:L-Phe & Periodic &  1.32$\pm$0.40 &  3676.57$\pm$407.47  \\
 L-Phe:L-Lys & Periodic &  1.89$\pm$2.14   &  2998.11$\pm$785.34     \\
 L-Phe & Periodic & $-$0.17$\pm$0.04   &  4316.33$\pm$948.98    \\
 L-Asp & Periodic & $-$0.20$\pm$2.37  &  2520.91$\pm$700.53      \\
	\hline
	\end{tabular}
\label{afdsasbbs}
\end{table}

\begin{table}[!tbp]
\centering
\caption{The Protenoids L-Glu:L-Phe:L-His, L-Glu:L-Phe,L-Phe:L-Lys, L-Asp, and L-Phe were illuminated with cold black light (695.8 lux) and cold white light (160.8 klux) for 30 minute on/30 minute off cycles, and the peak voltage and period were recorded.}
\begin{tabular}{|c|c|c|c|}
	\hline
	Proteinoid  & Light & Peaks & Periods      \\
       &  Condition  &  Mean  &  Mean     \\
    &  & (mV) & (sec)      \\
	\hline\hline
	
	L-Glu:L-Phe:L-His& Periodic & 6.89$\pm$0.28  &    3548$\pm$ 684.10  \\
 L-Glu:L-Phe & Periodic & 1.19$\pm$0.07   & 3625$\pm$668.24   \\
 L-Phe:L-Lys & Periodic &  $-$0.80$\pm$0.61   &  3673.14$\pm$397.16     \\
 L-Phe & Periodic &   $-$0.17$\pm$0.02 &  3620.62$\pm$856.44    \\
 L-Asp & Periodic & $-$10.07$\pm$4.05 & 3528.75$\pm$684.41     \\
	\hline
	\end{tabular}
\label{jkcbavljkbv}
\end{table}

\begin{table}[!tbp]
\centering
\caption{The Protenoids L-Glu:L-Phe:L-His, L-Glu:L-Phe:L-Lys, and L-Phe were illuminated with cold black light (695.8 lux) and cold white light (134.2 klux) for 30 minute on/30 minute off cycles, and the peak voltage and period were recorded.}
\begin{tabular}{|c|c|c|c|}
	\hline
	Proteinoid  & Light & Peaks & Periods      \\
       &  Condition  &  Mean  &  Mean     \\
    &  & (mV) & (sec)      \\
	\hline\hline
	
	L-Glu:L-Phe:L-His& Periodic & 7.57$\pm$0.91   &  1800.88$\pm$423.86    \\
 L-Glu:L-Phe & Periodic & 0.36$\pm$0.10 & 2820.77$\pm$647.68    \\
 L-Phe:L-Lys & Periodic & 10.21$\pm$0.78   &4484.83$\pm$495.45       \\
 L-Phe & Periodic & $-$0.08$\pm$0.04  &  3600$\pm$801.04     \\
 L-Asp & Periodic & $-$5.01$\pm$1.94  & 2309.31$\pm$554.36        \\
	\hline
	\end{tabular}
\label{lksvn;lsv}
\end{table}

%\bibliographystyle{plain}
%\bibliography{proteinoid_electrical_activity,mybibfile}

%\bibliography{proteinoid_electrical_activity,mybibfile}